\documentclass[aps,prd,twocolumn,showpacs,nofootinbib,amsmath,amssymb,amsfonts,showkeys]
{revtex4}

\usepackage{graphicx}
\usepackage{xcolor}
\usepackage{diagbox}

\begin{document}
	
	\title{Absorption effects in the expanding Universe: spectral transmittance \\ functions of the intergalactic medium for distant sources}
	\author{A. Yang$^1$, B. Novosyadlyj$^{1,2\,*}$, B. Melekh$^2$, G. Milinevsky$^{1,3,4}$\footnote{corresponding authors}}
    \affiliation{$^1$Jilin University, Qianjin Street 2699, Changchun, 130012, P.R.China,\\
    $^2$Ivan Franko National University of Lviv, Kyryla i Methodia str., 8, Lviv, 79005, Ukraine \\
    $^3$National Antarctic Scientific Center, 16 Taras Shevchenko Blvd., Kyiv 01601, Ukraine\\
    $^4$Main Astronomical Observatory of NASU, 27 Akademika Zabolotnoho Str., 03143, Kyiv, Ukraine}

	\date{\today}
	
\begin{abstract}
We construct two self-consistent analytic approximations to the neutral hydrogen fraction, $x_{\rm HI}(z)$, and the helium ionization fractions, $x_{\rm HeI}(z)$, $x_{\rm HeII}(z)$, and $x_{\rm HeIII}(z)$, that are consistent with current constraints inferred from quasar spectra, galaxy surveys, and CMB polarization measurements. These approximations describe observationally motivated early- and late-reionization scenarios. Using these histories, we analyse the formation of broad absorption troughs in the continuum spectra of high-redshift sources over $1\leq z_{\rm s}\leq15$. We assume that neutral hydrogen and helium in a homogeneous diffuse intergalactic medium reside predominantly in their ground states and absorb radiation through the Lyman-series lines and continua of HI, HeI, and HeII. We compute the wavelength-dependent optical depths for the first 39 Lyman-series lines of HI and HeII, the first 10 lines of HeI, and the corresponding continua, and use them to derive spectral transmittance functions, $T(\lambda;z_{\rm s})$. As illustrative applications, we apply them to toy-model continuum spectra of starless haloes and to model spectra of a low-metallicity dwarf galaxy at different redshifts. Spectral features in sources at $5\lesssim z_{\rm s}\lesssim7$ caused by intergalactic absorption are found to be particularly sensitive to the adopted hydrogen and helium ionization histories.
\end{abstract}
\pacs{95.36.+x, 98.80.-k}	 
\keywords{intergalactic medium, Gunn--Peterson effect, cosmological reionization, Cosmic Dawn, spectral transmittance}
\maketitle

\section{Introduction}

The first billion years of cosmic history represent a key period for addressing several outstanding problems in modern cosmology and astrophysics, including the formation of the first stars and galaxies and the origin of tensions in the determination of cosmological parameters. The epoch extending from cosmological recombination ($z\simeq1090$, $\sim$380 kyr) to the completion of reionization ($z\simeq6$, $\sim$1 Gyr) is characterized by an intergalactic medium dominated by neutral hydrogen and helium. This medium absorbs radiation from the earliest luminous sources over a wide range of wavelengths, strongly suppressing their direct detectability by optical and near-infrared observations. Knowledge of the reionization history of the intergalactic medium therefore provides crucial constraints on models of the first ionizing sources, while, conversely, the spectra of distant sources encode information about the intervening absorbers along the line of sight.

Recent high-redshift observations and cosmological simulations have substantially improved our understanding of the thermodynamic and ionization state of the intergalactic medium during the epoch of reionization and the early stages of galaxy formation \cite{Planck2020a,Planck2020b,Kageura2025,Gnedin2014,Puchwein2019}. These advances make it possible to confront analytical descriptions of intergalactic absorption with observationally constrained reionization histories and detailed radiative-transfer calculations.

Measurements of CMB temperature anisotropy and polarization constrain the ionization history of the Universe. In particular, measurements of large-scale CMB polarization by the \textit{Planck} mission \cite{Planck2020a,Planck2020b,Glazer2018} yield strong constraints on the redshift evolution of the free-electron fraction, $x_{\rm e}(z)$ at $z\le50$. Complementary information on the neutral hydrogen fraction, $x_{\rm HI}(z)$, is obtained from the analysis of Ly$\alpha$ hydrogen absorption and line asymmetries in the spectra of high-redshift sources, first discussed by Gunn \& Peterson \cite{Gunn1965} and Scheuer \cite{Scheuer1965} and subsequently explored using large ground-based and space-borne telescopes \cite{Fan2000,Fan2001,Becker2001,Fan2002,Pentericci2002,Becker2013,Inoue2014,Kamble2020,Bosman2022,Kageura2025}. A comprehensive compilation of current measurements of $x_{\rm HI}$ over the redshift range $5\lesssim z\lesssim14$ is presented in \cite{Kageura2025}. The fractions of neutral and ionised helium are discussed in \cite{MiraldaEscude2003,Furlanetto2006,Worseck2011,McQuinn2016,Makan2021} on the basis of direct and indirect observational evidence on spectral features of distant quasars and galaxies. On the basis of these works we deduce two self-consistent analytic approximations to the mean fractions $x_{\rm HI}(z)$, $x_{\rm HeI}(z)$, $x_{\rm HeII}(z)$, and $x_{\rm HeIII}(z)$ that represent ``early'' and ``late'' reionization scenarios and ionization histories up to $z=15$. Constructing these analytic ionization histories is the first objective of this work.

Although intergalactic absorption by hydrogen and helium has been extensively studied, most existing treatments are formulated either in terms of effective optical depths \cite{Madau1995,Meiksin2006} or rely on numerical radiative-transfer simulations. In this work we focus on an analytical description of these absorption processes, emphasizing their impact on observable spectra. The second goal of this paper is to analyse the formation of broad absorption troughs in the spectra of distant sources caused by absorption in the Lyman-series lines and continua of HI, HeI, and HeII, taking into account observationally constrained reionization histories and source redshifts in the range $1\le z\le15$. To illustrate these effects, we compute wavelength-dependent optical depths and apply them to toy continuum spectra of starless halos and to a model spectrum of a low-metallicity dwarf galaxy with a mass of $\sim10^9\,\mathrm{M_\odot}$.

Our approach is not intended to replace clumped absorber-based transmission prescriptions. It provides a complementary analytic baseline for a smoothly distributed IGM, allowing the effects of the mean hydrogen and helium ionization histories to be isolated. We explicitly compare our results with those of \cite{Madau1995} and \cite{Inoue2014} at $z\lesssim5$.

This paper is organized as follows. In Section~2 we derive the analytical approximations for two observationally motivated reionization histories of hydrogen and  helium based on the observational data sets. In Section~3 we compute the wavelength dependence of optical depths in the Lyman-series lines and continua of HI. Section~4 extends this analysis to absorption by HeI and HeII. In Section~5 we introduce spectral transmittance functions of the homogeneous intergalactic medium that are independent of the intrinsic source spectra. In Section 6 we illustrate these results using model spectra of starless haloes and a low-metallicity dwarf galaxy at different redshifts. Our conclusions are summarized in Section~7.

\section{Observational data on the reionization history}

To study absorption effects in the expanding Universe, observable from the Earth or relevant for computing the radiation energy density during the epoch of reionization,
it is necessary to specify the number density of absorbers along the line of sight. In this paper, we construct an analytic approximation to observational constraints on
the neutral hydrogen fraction, $x_{\rm HI}(z)$, as well as the helium fractions $x_{\rm HeI}(z)$, $x_{\rm HeII}(z)$, and  $x_{\rm HeIII}(z)$ based on measurements from several independent probes.

These include the large-scale CMB polarization data from the \textit{Planck} mission \cite{Planck2020b}, high-resolution spectra of the XQR-30 quasar sample obtained with
the Very Large Telescope, which provide evidence for a late end of reionization at $z\simeq5.3$ \cite{Bosman2022}, and James Webb Space Telescope spectra of approximately
600 galaxies, from which Kageura et al. \cite{Kageura2025} inferred $x_{\rm HI}$ at $z\simeq6$--14. The latter work also provides a comprehensive compilation of $x_{\rm HI}$  from the literature obtained using different methods (see their Fig.~11).  To approximate the mean neutral hydrogen fraction at $z<5$, we use Ly$\alpha$-forest transmission data from quasar spectra.

\begin{figure}
\includegraphics[width=0.49\textwidth]{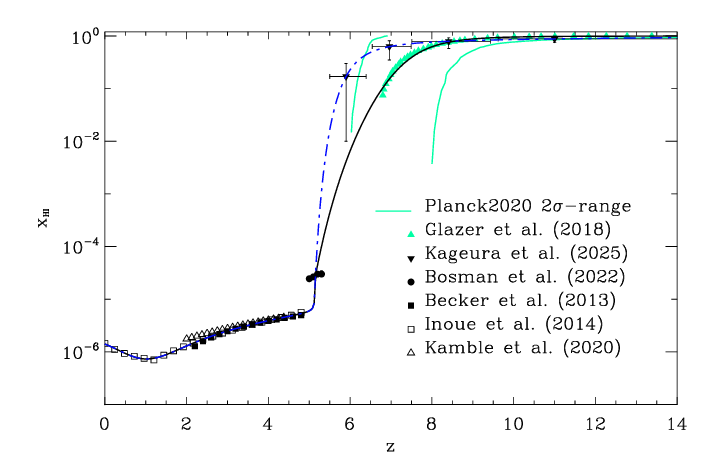}
\includegraphics[width=0.49\textwidth]{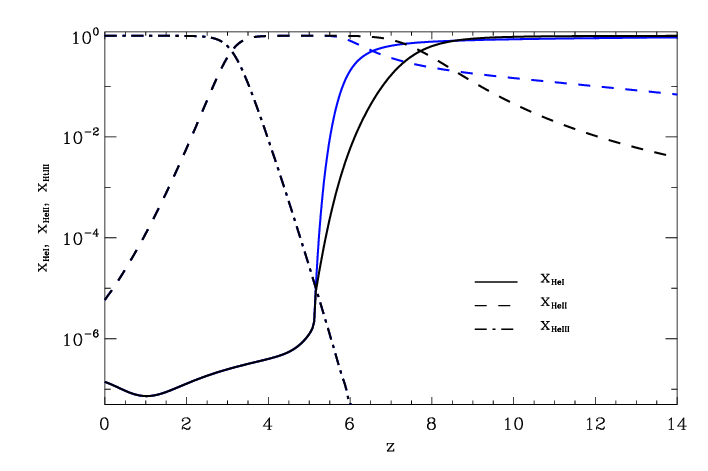}
\caption{Evolution of the neutral hydrogen fraction (top) and helium ionization fractions (bottom) for the two observationally motivated histories, KBBI (blue curves) and GBBI (black curves).}
\label{xHI-xHe}
\end{figure}

In this study, for $z>5$ we adopt the $x_{\rm HI}(z)$ estimates from \cite{Kageura2025,Bosman2022}, which are shown in the top panel of Fig.~\ref{xHI-xHe} as data points with error bars. The
$2\sigma$ range of the \textit{Planck} constraints is indicated by green lines, while
the median values inferred in \cite{Glazer2018} are shown by the green triangles\footnote{The
original CMB constraints are given for the free-electron fraction $x_{\rm e}(z)$.
Assuming $x_{\rm HeI}(z)=x_{\rm HI}(z)$ for $z\ge6$, we obtain $x_{\rm HI}=1-x_{\rm e}(z)/(1+f_{\rm He})$, where
$f_{\rm He}\equiv n_{\rm He}/n_{\rm H}$.}. For lower redshift, $z<5$, we use the Ly$\alpha$ transmission data from \cite{Becker2013,Inoue2014} to recover\footnote{Using the analytical form (\ref{tauLya}) for $\tau^{\rm HI}_{\rm Ly_\alpha}$, the analytical approximation $\tau_{\rm eff}(z)$ from \cite{Becker2013} (Fig.8, Tab. 3) and/or the analytical approximation $\tau^{\rm LAF}_{\rm 2}(z)$ given by eguation (21) in \cite{Inoue2014} one can obtain $x_{\rm HI}(z)$ for any cosmological model.} $x_{\rm HI}$. More recently, effective Ly$\alpha$ optical depths measured in high-redshift quasar spectra have also been used to constrain the photoionization state of the IGM \cite{Liu2025}. Other two curves represent two analytic approximations: blue dash-dotted line fits the data sets of \cite{Kageura2025,Bosman2022,Becker2013,Inoue2014} (KBBI or 1), and dark solid line fits the data sets of \cite{Glazer2018,Bosman2022,Becker2013,Inoue2014} (GBBI or 2). The analytic approximation consists of low- and high-redshift branches:
\begin{equation}
x_{\rm HI}(z)=
\begin{cases}
x_*\,\exp\!\left[
\dfrac{t\left(a_1+a_2 t+a_3 t^2+a_4 t^3\right)}
{1+b_1 t+b_2 t^2+b_3 t^3}
\right], & z\le z_*,\\[2.2ex]
\frac{1}{1+\exp\!\left(-\ln{\left(\frac{x_*}{1-x_*}\right)} - A\,s - B\,s^2 - C\,s^3\right)}, & z> z_*,
\end{cases}
\label{ai}
\end{equation}
where $t=z-z_*$, $s=\ln\!\left(1+\frac{z-z_*}{D}\right)$. Two branches of approximation join continuously at the point $(z_*,\,x_*)=(5.15,2.77125\cdot10^{-5})$ which corresponds to the averages of the four Bosman et al. (2022) \cite{Bosman2022} data points over $5.0\le z\le5.3$. The remaining parameters $a_1,\,a_2,\,a_3,\,a_4,\,b_1,\,b_2,\,b_3,\,A,\,B,\,C$ and $D$ are fitting parameters. Their values are determined by minimizing $\chi^2$ using the Levenberg--Marquardt method \cite{Press1992} and are listed in Tables~\ref{tab1}-\ref{tab1a}, where BBI denotes the data set of \cite{Bosman2022,Becker2013,Inoue2014}, KB those of \cite{Kageura2025,Bosman2022} and GB those of \cite{Glazer2018,Bosman2022}.
\begin{table*}
\centering
\caption{Best-fit parameters of the low-redshift branch of equation (\ref{ai}).}
\begin{tabular}{cccccccc}
\hline
\hline
Data set&$a_1$&$a_2$&$a_3$&$a_4$&$b_1$&$b_2$&$b_3$\\
\hline
BBI&159.1068&36.4547&-3.29829&-1.07885&-100.0000&-42.4907&-4.86781\\
\hline
\hline
\end{tabular}
\label{tab1}
\end{table*}\begin{table*}
\centering
\caption{Best-fit parameters of the high-redshift branch of equation (\ref{ai}).}
\begin{tabular}{cccccc}
\hline
\hline
No&Data set&$A$&$B$&$C$&$D$\\
\hline
1&KB&11.7083&-4.06938&0.523253&0.343817\\
2&GB&10.7053&-1.52525&0&1.14447\\
\hline
\hline
\end{tabular}
\label{tab1a}
\end{table*}

The approximation based on the data of \cite{Kageura2025} and \cite{Bosman2022} indicates that $x_{\rm HI}\approx0.5$ occurs at redshifts $z\approx6.5$. The approximation based on \cite{Glazer2018} and \cite{Bosman2022} yields the corresponding value at $z\approx7.7$, about 170 Myr earlier. We refer to the former case as a late reionization scenario and to the latter as an early reionization scenario.
In Table \ref{tab1b}, we present the fraction of neutral hydrogen for both approximations at specific redshifts $z\in1-15$ used in the following sections.
\begin{table*}
\centering
\caption{Fraction of neutral hydrogen $x_{\rm HI}$ at specific redshifts $z\in1-15$ for approximation KBBI (1) and GBBI (2).}
\begin{tabular}{cccccccccccc}
\hline
\hline
Model/$z$&1&2&3&4&$5$&$6$&$7$& $8$&$10$&$12$&$15$\\
\hline
1&$7.3\cdot10^{-7}$&$1.3\cdot10^{-6}$&$2.5\cdot10^{-6}$&$3.9\cdot10^{-6}$&$6.0\cdot10^{-6}$&$0.23$&$0.64$&$0.77$&$0.85$&$0.90$&$0.94$ \\
\hline
2&$7.3\cdot10^{-7}$&$1.3\cdot10^{-6}$&$2.5\cdot10^{-6}$&$3.9\cdot10^{-6}$&$6.0\cdot10^{-6}$&$6.6\cdot10^{-3}$&$0.17$&$0.62$&$0.95$&$0.990$&$0.997$ \\
\hline
\hline
\end{tabular}
\label{tab1b}
\end{table*}
 
\begin{table*}
\centering
\caption{Optical depth for Thomson scattering to sources at specific redshifts $z_{\rm s}\in1-15$ for approximation 1 (KBBI) and 2 (GBBI).}
\begin{tabular}{cccccccccccc|c}
\hline
\hline
Model/$z$&1&2&3&4&$5$&$6$&$7$& $8$&$10$&$12$&$15$&$50$\\
\hline
1&0.0034&0.0087&0.015&0.022&$0.030$&$0.039$&$0.045$&$0.048$&$0.0523$&$0.0555$&$0.0589$&$0.0634$ \\
\hline
2&0.0034&0.0087&0.015&0.022&$0.030$&$0.040$&$0.049$&$0.056$&$0.0599$&$0.0604$&$0.0607$&$0.0610$ \\
\hline
\hline
\end{tabular}
\label{tab1c}
\end{table*}

At $z\gtrsim5$, the observationally inferred neutral hydrogen fraction should be interpreted with some caution. During the epoch of Reionization, the intergalactic medium is expected to be highly inhomogeneous, consisting of ionized regions surrounding sources and still-neutral patches that have not yet been reionized. Consequently, estimates of $x_{\rm HI}$ obtained from damping-wing analyses, Ly$\alpha$ transmission statistics, or related probes generally constrain a global or volume-averaged neutral fraction rather than the residual neutral fraction inside the already ionized regions. The latter is model-dependent and is typically much smaller, but it is this residual neutral component, together with self-shielded systems and partially neutral structures, that determines the detailed line-of-sight transmission through ionized regions. In the present work, the observationally constrained $x_{\rm HI}(z)$ is therefore used as an effective homogeneous representation of the global reionization history. The resulting optical depths and transmittance functions at $z\gtrsim5$ should accordingly be regarded as a homogeneous limiting case that isolates the dependence on the mean ionization history, rather than as a detailed model of patchy reionization or of individual sight lines.

Helium provides an additional contribution to intergalactic absorption and to the free-electron budget relevant for Thomson scattering. During the early stages of reionization, the first ionization of helium (HeI$\rightarrow$HeII) is expected to proceed nearly concurrently with hydrogen reionization, driven primarily by stellar-dominated ionizing backgrounds \cite{MiraldaEscude2003,Furlanetto2006}. At later times, helium undergoes a second ionization (HeII$\rightarrow$HeIII), commonly associated with the emergence of hard UV radiation from quasars and observationally constrained by HeII Ly$\alpha$ absorption measurements \cite{Worseck2011,Makan2021}. The thermal evolution inferred from Ly$\alpha$ forest absorbers provides complementary constraints on this history and is consistent with the completion of HeII reionization around $z\sim3$ \cite{Yang2023}.

In this work we describe the ionization state of helium using phenomenological ionization fractions that are consistent with observational constraints while remaining analytically tractable. At high redshift we assume that the first ionization of helium proceeds concurrently with hydrogen reionization, such that $x_{\mathrm{HeI}} \simeq x_{\mathrm{HI}}$. At lower redshift, where the diffuse intergalactic medium is highly ionized, we parametrize the residual neutral helium fraction by a proportionality $x_{\mathrm{HeI}}=\eta\,x_{\mathrm{HI}}$, where $\eta\in[0.1,1]$ brackets the uncertainty associated with the ultraviolet background.

To connect these two regimes smoothly, we introduce a transition function
\begin{equation}
W(z)=\frac12\left[1+\tanh\!\left(\frac{z-z_t}{\Delta z}\right)\right],
\end{equation}
with $z_t=5.5$ and $\Delta z=0.5$, and define
\begin{equation}
x_{\mathrm{HeI}}(z)=x_{\mathrm{HI}}(z)\left[\eta+(1-\eta)\,W(z)\right].
\end{equation}
This construction ensures a continuous and differentiable ionization history,
recovering $x_{\mathrm{HeI}}=x_{\mathrm{HI}}$ at $z\gtrsim6$ and
$x_{\mathrm{HeI}}=\eta\,x_{\mathrm{HI}}$ at $z\lesssim5$.

The doubly ionized helium fraction is modeled independently in order to capture
the second ionization of helium (HeII$\rightarrow$HeIII), which
is commonly associated with hard ultraviolet radiation from quasars at Cosmic Noon and is
observationally constrained by HeII Ly$\alpha$ absorption measurements.
We adopt a smooth transition of the form
\begin{equation}
x_{\mathrm{HeIII}}(z)=\frac12\left[1+\tanh\!\left(\frac{z_{\mathrm{He}}-z}
{\Delta z_{\mathrm{He}}}\right)\right],
\end{equation}
where $z_{\mathrm{He}}$ denotes the characteristic redshift of helium reionization
and $\Delta z_{\mathrm{He}}$ its duration.

Observations of HeII Ly$\alpha$ absorption indicate that helium
reionization is largely completed by $z\simeq2.7$ and proceeds over a relatively
extended redshift interval, motivating typical parameter values
$z_{\mathrm{He}}\simeq2.8$--$3.5$ and $\Delta z_{\mathrm{He}}\simeq0.4$--$0.6$
\cite{Worseck2011,McQuinn2016,Makan2021}. The singly ionized helium fraction then
follows from the normalization condition
\begin{equation}
x_{\mathrm{HeII}}(z)=1-x_{\mathrm{HeI}}(z)-x_{\mathrm{HeIII}}(z).
\end{equation}

At low redshift, the intergalactic medium is expected to approach a photoionization equilibrium state in which helium is almost fully doubly ionized. Ultraviolet background models calibrated to Ly$\alpha$ and HeII Ly$\alpha$ observations predict that the diffuse IGM at low redshifts is dominated by HeIII, with a small residual fraction of singly ionized helium and a negligible neutral helium abundance, $x_{\rm HeI}:x_{\rm HeII}:x_{\rm HeIII}\sim10^{-6}:10^{-3}:1$ \cite{HaardtMadau2012,Shull2012,Shull2015,McQuinn2016,FaucherGiguere2020}. This asymptotic behavior provides a useful reference point for assessing the physical plausibility of phenomenological helium ionization histories at intermediate redshifts. 

The redshift dependence of the helium ionization fractions resulting from the above
prescriptions is shown in the bottom panel of Fig.~\ref{xHI-xHe} for the two
$x_{\rm HI}$ approximations displayed in the top panel. The calculations are carried
out for $\eta=0.1$, $z_{\rm He}=3.1$, and $\Delta z_{\rm He}=0.4$. Variations of these
parameters within the ranges discussed above do not qualitatively alter the resulting
ionization histories. In this phenomenological model, the asymptotic relation
$x_{\rm HeI}:x_{\rm HeII}:x_{\rm HeIII}\sim10^{-6}:10^{-3}:1$ is already reached at
$z\sim1$--2.

Our approximations can be used to compute the redshift dependence of the free-electron fraction, $x_e(z)\equiv n_e(z)/n_{\rm H}(z)$, where $n_{\rm H}(z)=n_{\rm H}(0)(1+z)^3$, $n_e(z)=n_{\rm H}(0)(1-x_{\rm HI}(z))(1+z)^3+n_{\rm He}(0)(x_{\rm HeII}(z)+2x_{\rm HeIII}(z))(1+z)^3$. This allows us to check the consistency of the resulting ionization histories by comparison of the Thomson optical depth  
\begin{equation}
 \tau_{\rm Th}(z_{\rm s})=n_{\rm H}(0)c\sigma_{\rm T}\int_0^{z_{\rm s}}dz\,\frac{x_{\rm e}(z)(1+z)^2}{H_0\sqrt{\Omega_{\rm m}(1+z)^3+\Omega_{\Lambda}}},
 \label{tau}
\end{equation}
($c$ is the speed of light, $\sigma_{\rm T}$ is the Thomson scattering cross-section) with the Planck constraint \cite{Planck2020a} (Tab. 7). For $H_0=67.36$ km/s/Mpc, $\Omega_{\rm m}=0.3153$, $\Omega_{\rm b}=0.0493$ and $\Omega_{\Lambda}=0.6847$ we have $n_{\rm H}(0)=1.897\cdot10^{-7}$ cm$^{-3}$, $n_{\rm He}(0)=1.536\cdot10^{-8}$ cm$^{-3}$. The results are presented in Table \ref{tab1c}. We find that both estimates of $\tau_{\rm Th}$ at $z=15$ lie within the $1\sigma$ range of the value inferred by the Planck Collaboration for the $\Lambda$CDM model with the
same cosmological parameters, $\tau=0.0544\pm0.0073$ \cite{Planck2020a}. However, this
Planck value is quoted for integration up to $z=50$. For this reason, we also report
$\tau_{\rm Th}$ for $z_{\rm s}=50$. In this case, approximation~1 yields $\tau_{\rm Th}$ which lies marginally above the Planck 2018 $1\sigma$ upper limit, whereas approximation~2 remains within them. It should be noted that in the fitting procedure of the KB data we have used the asymptotic prior at $z = 40$ $x_{\rm HI}=0.9998$. Without it, $\tau_{\rm Th}$ goes far beyond the Planck 2018 limit. This comparison is not fully independent, because CMB information enters the construction of the GBBI history and motivates the high-redshift asymptotic prior adopted for the KBBI history.

These ionization histories provide the basis for computing the wavelength-dependent optical depths discussed in the following sections. 

\section{Optical depth of the diffuse intergalactic gas in hydrogen lines and continuum}

\begin{figure}
\includegraphics[width=0.49\textwidth]{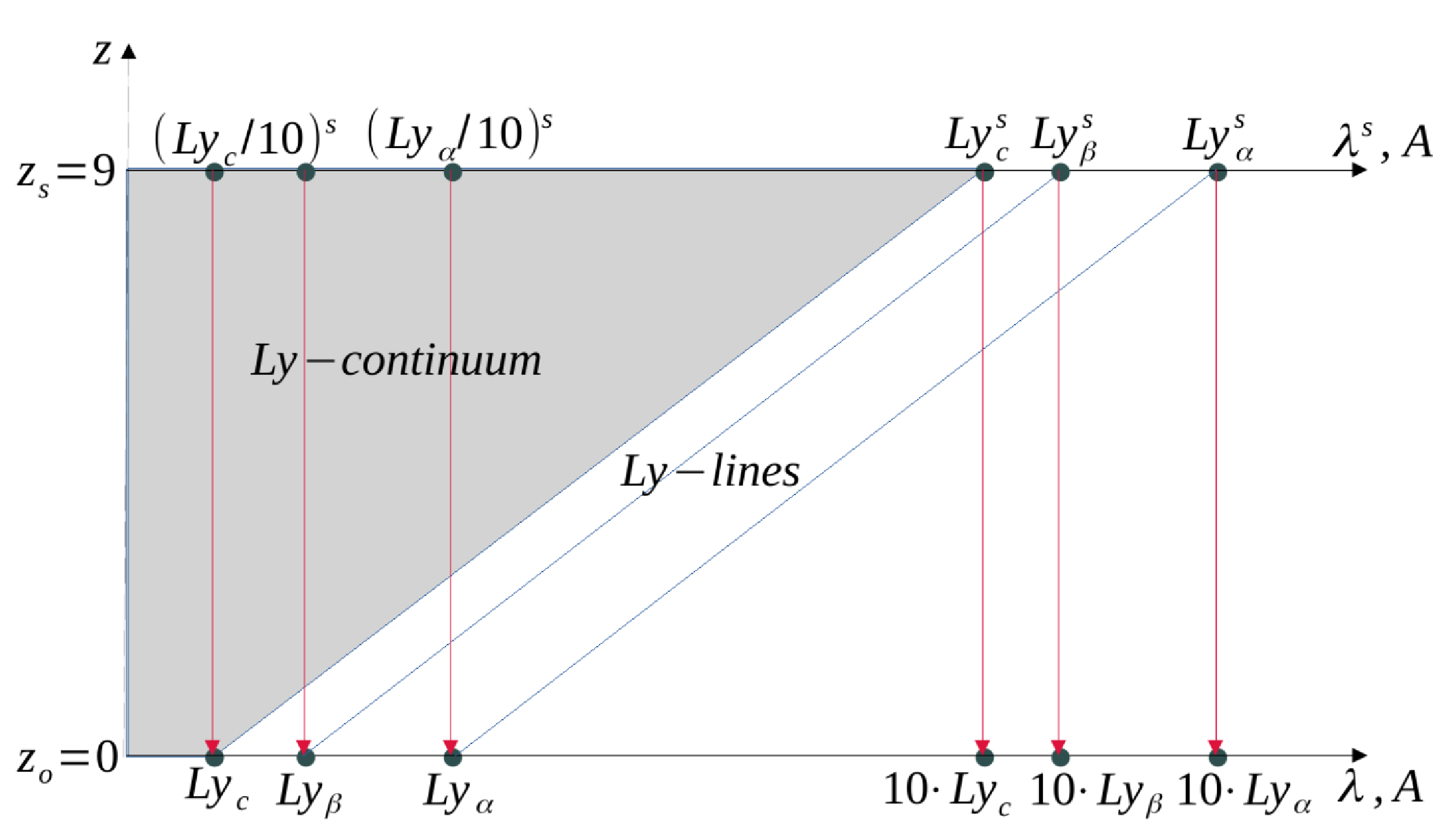}
\caption{Schematic representation of photon absorption along the line of sight from a source at $z_{\rm s}=9$.}
\label{scheme}
\end{figure}

Since the goal of this paper is to analyse the transmission of continuum spectra
from the first sources of light during the Cosmic Dawn and Reionization epochs, we do
not require detailed knowledge of the intrinsic source spectra. It is sufficient to
assume a generic spectral energy distribution, account for cosmological redshift,
identify the relevant absorbers and their spectral features, and integrate the
radiative transfer equation for the specific intensity $I_\nu$ at frequency $\nu=c/\lambda$ along the line of sight.
\begin{table*}
\centering
\caption{The wavelengths $\lambda^{\rm HI}_{\rm 1-n}$ ({\AA}) and oscillator strengths $f^{\rm HI}_{\rm 1-n}$ for the first 39 hydrogen lines of Lyman series \cite{Wiese2009}.}
\begin{tabular}{ccc|ccc|ccc}
\hline
\hline
1-n&$\lambda^{\rm HI}_{\rm 1-n}$&$f^{\rm HI}_{\rm 1-n}$&1-n&$\lambda^{\rm HI}_{\rm 1-n}$&$f^{\rm HI}_{\rm 1-n}$&1-n&$\lambda^{\rm HI}_{\rm 1-n}$&$f^{\rm HI}_{\rm 1-n}$ \\
\hline
1-2& 1215.67& 0.41641&1-15& 915.821& 0.00046886&1-28& 912.916& 7.1476E-05  \\
1-3& 1025.720& 0.079142&1-16& 915.327& 0.00038577&1-29& 912.837& 6.4319E-05 \\
1-4& 972.537& 0.029006&1-17& 914.917& 0.00032124&1-30& 912.765& 5.8087E-05 \\
1-5& 949.743& 0.013945&1-18& 914.574& 0.00027035&1-31& 912.701& 5.2635E-05 \\
1-6& 937.803& 0.0078035&1-19& 914.284& 0.000229671&1-32& 912.642& 4.7845E-05\\
1-7& 930.748& 0.0048164&1-20& 914.036& 0.00019677&1-33& 912.589& 4.3619E-05\\
1-8& 926.223& 0.0031850&1-21& 913.823& 0.00016987&1-34& 912.541& 3.9877E-05\\
1-9& 923.148& 0.0022172&1-22& 913.639& 0.00014767&1-35& 912.496& 3.6551E-05\\
1-10& 920.961& 0.0016062&1-23& 913.478& 0.00012917&1-36& 912.455& 3.3585E-05\\
1-11& 919.349& 0.0012011&1-24& 913.337& 0.00011364&1-37& 912.418& 3.0931E-05\\
1-12& 918.127& 0.0009219&1-25& 913.212& 0.00010051&1-38& 912.383& 2.8550E-05\\
1-13& 917.178& 0.0007231&1-26& 913.102& 8.9321E-05&1-39& 912.351& 2.6407E-05\\
1-14& 916.427& 0.00057769&1-27& 913.004& 7.9736E-05&1-40& 912.321& 2.4474E-05\\
\hline
\hline
\end{tabular}
\label{tab2}
\end{table*}
\begin{figure*}
\includegraphics[width=0.49\textwidth]{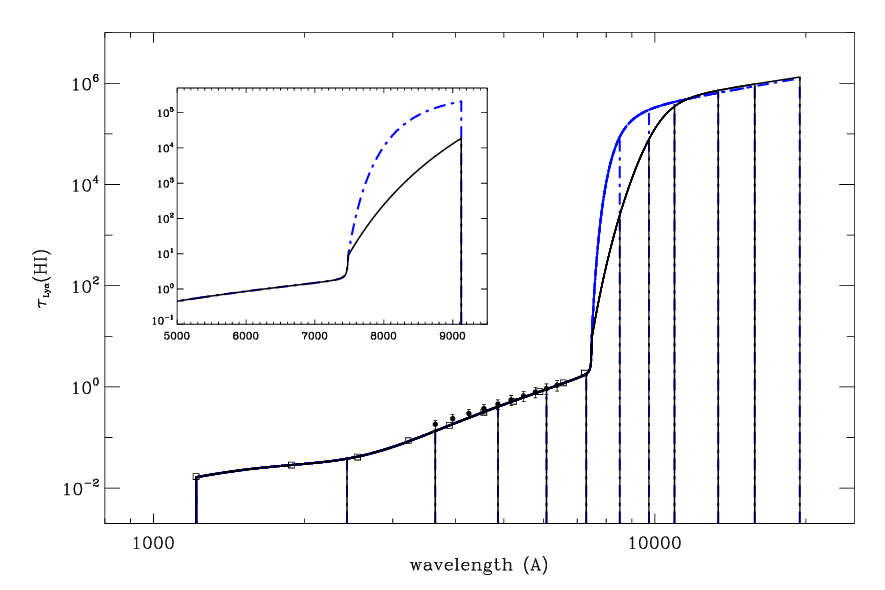}
\includegraphics[width=0.49\textwidth]{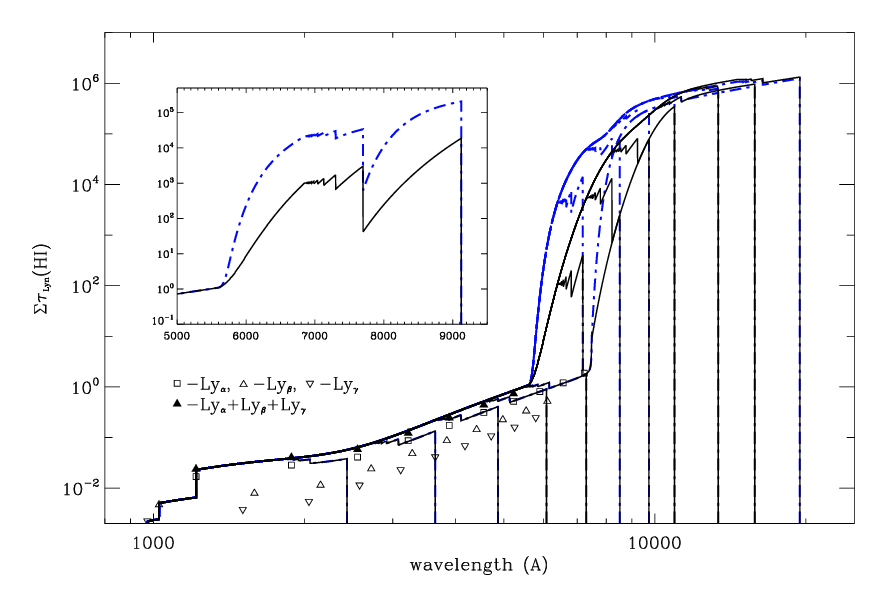}
\caption{Optical depth in the Ly$\alpha$ line (left panel) and superposition of the first 39 lines of Lyman series of HI (right panel) for sources at redshifts $z=1$, 2, 3, 4, 5, 6, 7, 8, 10, 12, 15 (from left to right). Dash-dotted blue lines show the optical depths for $x_{\rm HI}$-approximation 1, solid lines for approximation 2. The linear–logarithmic insets show the wavelength dependence of the optical depth for a source at $z=6.5$.}
\label{tauLyH}
\end{figure*}

In \cite{Yang2025}, we examined cosmological redshifting of source spectra in the absence of intervening matter. Here we extend that treatment by including absorption in the expanding intergalactic medium. We assume that hydrogen and helium atoms in the intergalactic medium reside predominantly in their ground states, since at $z\le14$ both the atomic number density and the density of energetic photons are too low to produce a statistically significant population of excited levels.
 
We analyse here the hydrogen (HI) and helium (HeI and HeII) absorption in the lines of Lyman series and Lyman-continuum. Figure \ref{scheme} illustrates the $\lambda-z$ absorption ranges by hydrogen for a source at $z_{\rm s}=9$. The upper $\lambda^{\rm s}$-axis shows wavelengths in the source rest frame, whereas the lower axis shows the corresponding wavelengths observed at $z_{\rm o}=0$. The radiation with $\lambda>\lambda_{\alpha}(1+z_{\rm s})$ is not absorbed by intergalactic medium. The radiation with  $\lambda_{\rm c}(1+z_{\rm s})<\lambda\le \lambda_{\alpha}(1+z_{\rm s})$ is absorbed in the lines of Lyman series at redshift where line $\lambda^{\rm s}-\lambda$ crosses the line Ly$^{\rm s}_{\rm n}$-Ly$_{\rm n}$, where ${\rm n}$ denotes $\alpha$, $\beta$, $\gamma$, ... . We refer to this region as the Lyman lines absorption domain in $\lambda-z$ space. The radiation with $\lambda\le \lambda_{\rm c}/(1+z_{\rm s})$ is absorbed in each redshift $z\in0-z_{\rm s}$ at corresponding local wavelength $\lambda=\lambda_{\rm c}(1+z)/(1+z_{\rm s})$. This $\lambda-z$ absorption range is shaded grey in Fig. \ref{scheme} and is referred to as the Lyman continuum absorption domain. The radiation with $\lambda_{\rm c} <\lambda<\lambda_{\rm c}(1+z_{\rm s})$ is absorbed in both $\lambda-z$ absorption domains at corresponding $0\le z\le z_{\rm s}$.

The expression for optical depth in the Ly$\alpha$ line for cosmological sources was obtained by Gunn \& Peterson \cite{Gunn1965}. We generalize it for all lines considered here and for any set of cosmological parameters.

\begin{figure*}
\includegraphics[width=0.49\textwidth]{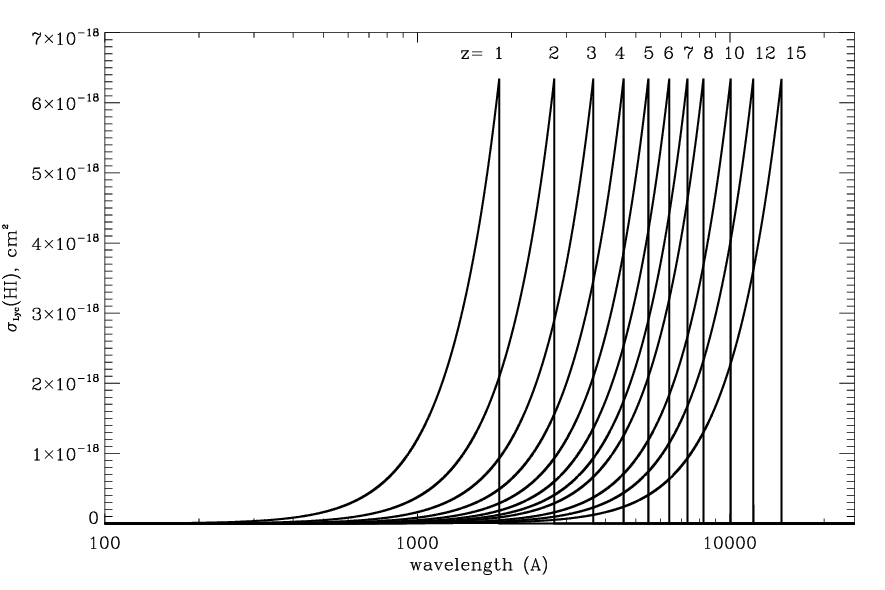}
\includegraphics[width=0.49\textwidth]{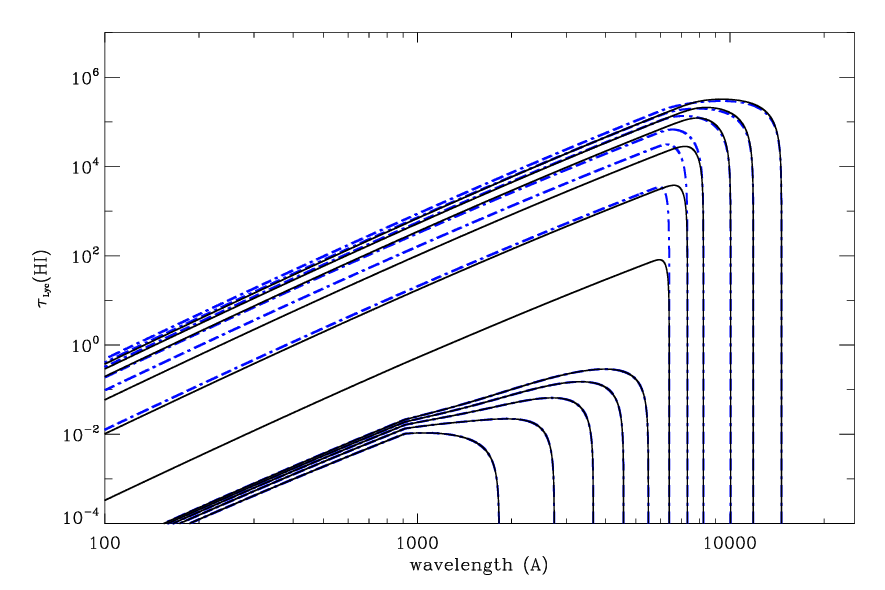}
\caption{The wavelength dependences of the effective cross-section of absorption in the Lyman-continuum in the rest frame of Earth observer transferred from the absorber reference system at different redshifts (left) and optical depth for distant sources at $z_{\rm s}\in1-15$ (right) for both approximations of $x_{\rm HI}$ (1 - dash-dotted blue lines, 2 - solid dark lines). }
\label{ftauLyc}
\end{figure*}

\begin{figure*}
\includegraphics[width=0.49\textwidth]{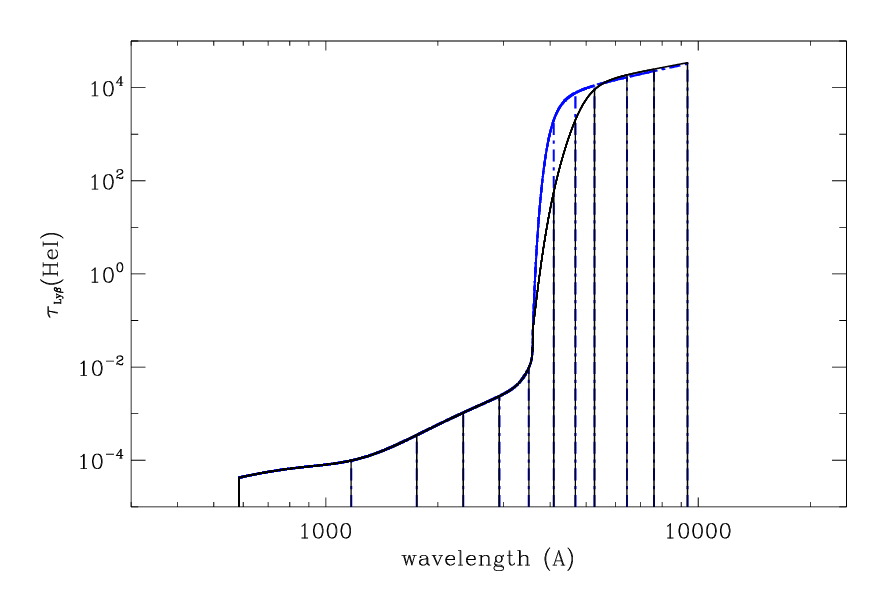}
\includegraphics[width=0.49\textwidth]{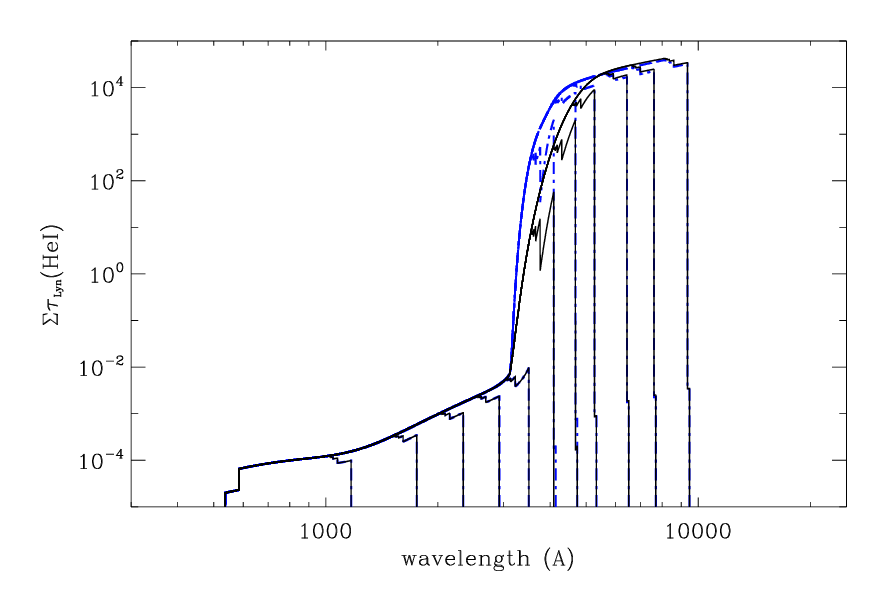}
\caption{Optical depth in the $Ly_\beta$ line (left panel) and superposition of the first 10 lines of Lyman series of HeI (right panel) for the sources at $z_{\rm s}\in1-15$. Blue dash-dotted lines are shown for approximation 1, and dark solid lines for approximation 2.}
\label{tauLyHe}
\end{figure*}

The effective cross-section of absorption in the Ly$\alpha$ line is
\begin{equation}
 \sigma^{\rm HI}_\alpha(\nu,z)=\alpha_{fs}\frac{h_{\rm P}}{2m_e}f^{\rm HI}_\alpha\Phi_\alpha(\nu;b,z),
 \label{sigma}
\end{equation}
where $\alpha_{\rm fs}$ is the fine-structure constant, $h_{\rm P}$ is Planck’s
constant, $m_e$ is the electron mass, $f^{\rm HI}_\alpha$ is the oscillator
strength of the transition, and $\Phi_\alpha$ denotes the Ly$\alpha$ line profile. The numerical values of the effective cross-sections here and below are given in cm$^2$. The line profile can be approximated by a Voigt profile \cite{Voigt1913,Irsic2018}:
\begin{eqnarray}
&&\Phi_\alpha(\nu,z)\approx\frac{1}{b_\nu\sqrt{\pi}}\left[\frac{A}{\sqrt{\pi}(A^2+B^2)}+e^{-B^2}\right], \nonumber \\
&&A=\frac{A_{21}}{4\pi b_\nu}, \quad B=\frac{\nu(1+z)/(1+z_{\rm o})-\nu_\alpha}{b_\nu},
\end{eqnarray}
where $b_\nu=\frac{b\nu_\alpha}{c}$ is the frequency width of the line and $b=\sqrt{\frac{2k_BT_b(z)}{m_H}+b^2_{turb}}$ is rms velocity caused by thermal and bulk turbulent motions of the hydrogen atoms in the intergalactic medium. Observations support the thermal broadening of lines with $T_{\rm b}\lesssim2\cdot10^4$ K at the end of reionization. Therefore, the redshift interval $\Delta z$ over which the incoming continuum radiation is absorbed by local neutral hydrogen does not exceed $\sim5\cdot10^{-4}$ and is very narrow. Since the integral $\int_{-\infty}^{+\infty}\Phi_\alpha(\nu,z)d\nu=1$, the expression for optical depth in this line in the $\Lambda$CDM model has analytical form as follows
\begin{eqnarray}
&&\tau^{\rm HI}_{\rm Ly_\alpha} (\lambda,z)=\frac{c}{H_0} \int ^{z_{\rm s}}_{z_{\rm o}}\frac{n_{\rm HI}(z)\sigma^{\rm HI}_\alpha(\lambda,z)dz}{(1+z)\cdot\sqrt{\Omega_{\rm m}\cdot(1+z)^3+\Omega_\Lambda}}\nonumber \\
&&= 9.2\cdot10^{10}\lambda^{\rm HI}_\alpha f^{\rm HI}_\alpha (1-Y_{\rm p})\frac{\Omega_{\rm b}h(1+z)^3x_{\rm HI}(z)}{\sqrt{\Omega_{\rm m}(1+z)^3+\Omega_\Lambda}},
\label{tauLya}
\end{eqnarray}
which for $z\gg1$ and known atomic constants becomes the same $\tau^{\rm HI}_{\rm GP}$ as in \cite{Becker2001,Fan2002}. In the last part of expression, the redshift along of line of sight is related with wavelength as $z=(1+z_{\rm s})\lambda^{\rm HI}_{\alpha}/\lambda-1$, where $\lambda^{\rm HI}_\alpha\le \lambda \le\lambda^{\rm HI}_{\alpha}(1+z_{\rm s})$, $z_{\rm s}$ is the redshift of source and $z_{\rm o}$ is the redshift of observer. In this paper all computations are carried out for $z_{\rm o}=0$.

Equation (\ref{tauLya}) is readily generalized to Ly$\beta$, Ly$\gamma$, and higher Lyman-series transitions by substituting the corresponding wavelengths and oscillator strengths 
{\small \begin{equation}
\tau^{\rm HI}_{\rm Ly_n}(z)=9.2\cdot10^{10}\lambda^{\rm HI}_{\rm 1-n} f^{\rm HI}_{\rm 1-n} (1-Y_{\rm p})\frac{\Omega_{\rm b}h(1+z)^3x_{\rm HI}(z)}{\sqrt{\Omega_{\rm m}(1+z)^3+\Omega_\Lambda}},
\label{tauLyn}
\end{equation} } where $z=(1+z_{\rm s})\lambda^{\rm HI}_{\rm 1-n}/\lambda-1$. In our computations we use the first 39 lines of Lyman series for which the values of wavelengths $\lambda^{\rm HI}_{\rm 1-n}$ and oscillator strengths $f^{\rm HI}_{\rm 1-n}$ are taken from \cite{Wiese2009}. Their values are listed in  Table \ref{tab2}. A cursory inspection of these quantities suggests that only a few of the lowest energy levels of the Lyman series are important in terms of influencing the observed spectrum of distant light sources. Below we demonstrate which transitions must be included for accurate calculations.

In the left panel of Fig. \ref{tauLyH} we present the dependence of $\tau^{\rm HI}_{\rm Ly_\alpha}$ on the wavelength for sources at different redshifts for both approximations of $x_{\rm HI}(z)$, presented in Fig. \ref{xHI-xHe} (dash-dotted blue lines for approximation 1, solid dark line for approximation 2). Figure~\ref{tauLyH} shows  that for sources at $z\le4$ the optical depth $\tau^{\rm HI}_{\rm Ly_\alpha}\le1$ and exceeds unity for sources at $z>5$ for both reionization histories. We also note that optical depths produced by Ly$\alpha$ absorption for sources at $5\lesssim z\lesssim 8$ are large for both reionization histories but noticeably lower for history 2 than for history 1.

Equations (\ref{tauLya})-(\ref{tauLyn}) show that
\begin{equation}
\tau^{\rm HI}_{\rm Ly_n}(\lambda;z)\propto\tau^{\rm HI}_{\rm Ly_\alpha}\frac{\lambda^{\rm HI}_{\rm 1-n}f^{\rm HI}_{\rm 1-n}}{\lambda^{\rm HI}_\alpha f^{\rm HI}_\alpha},
\label{taun-a}
\end{equation}
thus, $\tau^{\rm HI}_{\rm Ly_n}$ decreases with $n$ at any $\lambda$ and $z$. This allows us to assess the importance of taking into account other lines of Lyman series. At $z\approx6$, the optical depth falls below unity only for transitions from the
ground state to levels with $n\gtrsim25$. At higher redshifts, all 39 Lyman-series
lines considered here become optically thick.
In the right panel of Fig. \ref{tauLyH} we show the optical depth of the first 39 lines of Lyman series $\tau^{\rm HI}_{\rm Ly}=\sum_2^{40}\tau^{\rm HI}_{\rm Ly_n}$ for the sources at different redshifts. The linear–logarithmic insets in Fig. \ref{tauLyH} show the wavelength dependence of the optical depth for a source at $z=6.5$. The higher Lyman-series lines substantially extend the absorption range towards shorter wavelengths. 

The photoionization cross-section in the Lyman continuum can be approximated by the analytic fit of Verner et al. \cite{Verner1996}:
{\footnotesize
\begin{align}
 \hskip-0.8cm\sigma^{\rm HI}_{\rm c} (\lambda,z)=\left\{
\begin{aligned} 
 &5.475\cdot10^{-14}\frac{\left(\frac{E}{E_0}-1\right)^2\left(\frac{E}{E_0}\right)^{-4.0185}}{\left(1+0.1744\sqrt{\frac{E}{E_0}}\right)^{2.963}},&\lambda\frac{1+z_{\rm o}}{1+z}\le \lambda_{\rm pi}\\
 &\hskip3.5cm0, &\lambda\frac{1+z_{\rm o}}{1+z}> \lambda_{\rm pi}
\end{aligned}
\right. , \label{sigma_c}
\end{align}}
where $\lambda_{\rm pi}=911.75$~{\AA} is wavelength corresponding to the potential of ionization of atomic hydrogen $E_{\rm pi}=13.6\,{\rm eV}$, and $E_0=0.4298\,{\rm eV}$ is a fitting parameter. The wavelength dependences of the effective cross-section (\ref{sigma_c}) in the rest frame of Earth observer transformed from the absorber rest frame to the observer frame at different redshifts by relation $\lambda^{\rm o}_{\rm pi}=\lambda_{\rm pi}(1+z)/(1+z_{\rm o})$ are shown in the left panel of Fig. \ref{ftauLyc}. 

The optical depth in this case is integral
\begin{equation}
\tau^{\rm HI}_{\rm c} (\lambda,z_{\rm s})=\xi\frac{cn^0_{\rm H}}{H_0}\int^{z_{\rm s}}_{z_{\rm o}}\frac{x_{\rm HI}(z) (1+z)^2 \sigma^{\rm HI}_{\rm c}(\lambda,z)dz}{\sqrt{\Omega_m(1+z)^3+\Omega_\Lambda}}, 
\label{tauLyc}
\end{equation}
where $n^0_{\rm H}=(1-Y_{\rm P})\frac{\Omega_{\rm b}}{m_{\rm H}}\frac{3H_0^2}{8\pi G}$, $\xi$ is fitting factor which is 1 in the most computations in this paper.    
The wavelength dependence of $\tau^{\rm HI}_{\rm c}$ for sources at different redshift for both approximations of $x_{\rm HI}$ is presented in the right panel of Fig. \ref{ftauLyc}. It shows that near the ionization threshold in the observer rest frame the optical depth $\tau^{\rm HI}_{\rm c}(\lambda\lesssim\lambda^{\rm HI}_{\rm pi}(1+z_{\rm s}))\lesssim1$ for $z_{\rm s}\lesssim5$ at all wavelengths, and is >1 for $z_{\rm s}\gtrsim5$. We also note significantly lower optical depths for sources at $z\approx6-8$ for the 2nd approximation of $x_{\rm HI}$ in comparison with the 1st one. Consequently, spectral features below about $\lambda\lesssim2000$~{\AA} also differ between the two reionization histories.  

\section{Optical depth of the diffuse intergalactic gas in the helium lines and continuum}
\begin{figure*}
\includegraphics[width=0.49\textwidth]{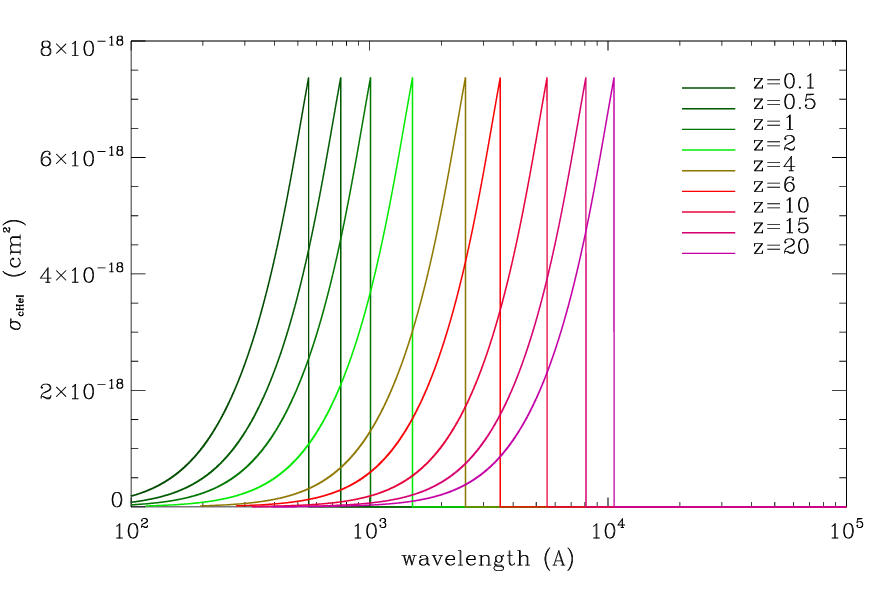}
\includegraphics[width=0.49\textwidth]{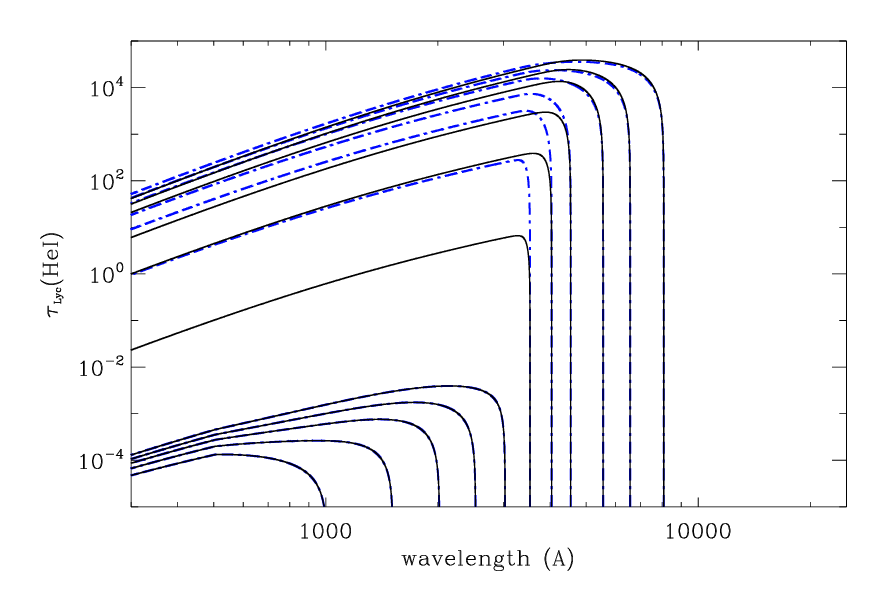}
\caption{The wavelength dependences of the effective cross-section of absorption in the HeI Lyman-continuum in the rest frame of Earth observer transferred from the absorber reference system at different redshifts (left) and optical depth in the HeI Lyman-continuum for distant sources at $z_{\rm s}\in1-15$ (right). }
\label{sigma-tauHeLyc}
\end{figure*}
\begin{figure*}
\includegraphics[width=0.49\textwidth]{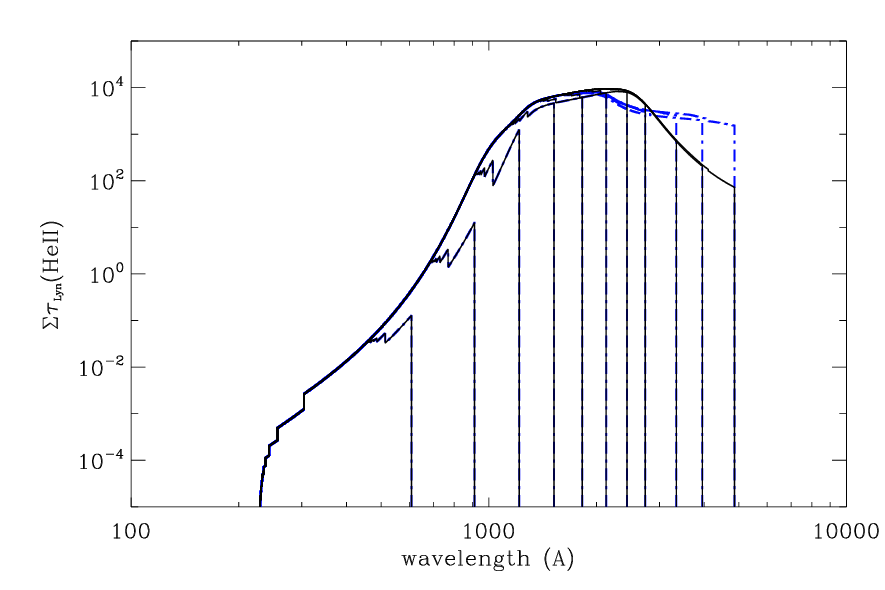}
\includegraphics[width=0.49\textwidth]{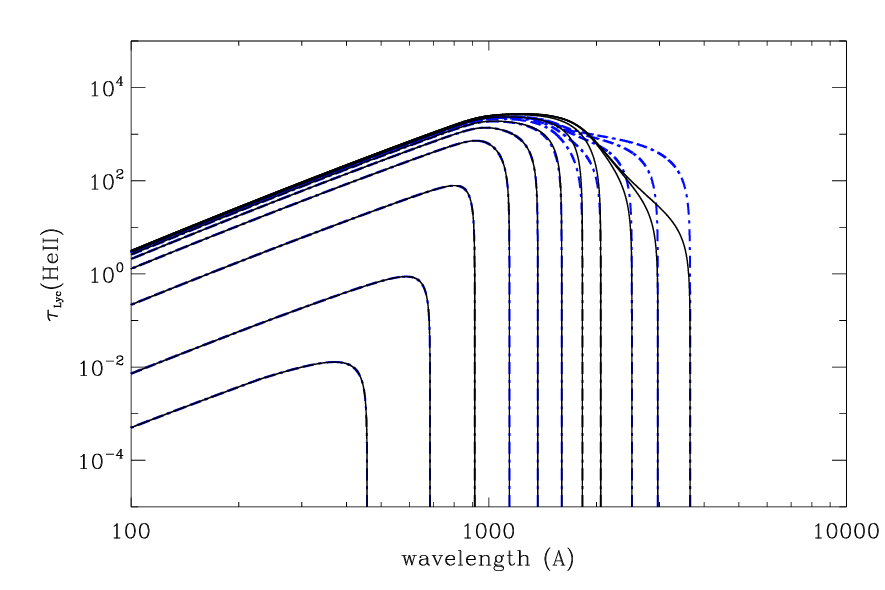}
\caption{The optical depths in the HeII lines of Lyman series (superposition of 39 lines, left panel) and in Lyman-continuum (right panel) for distant sources at $z_{\rm s}\in1-15$. }
\label{tauHeII}
\end{figure*}

The helium reionization history adopted here, shown in the bottom panel of Fig.~\ref{xHI-xHe}, indicates that at $z\gtrsim6$ the neutral helium fraction $x_{\rm HeI}(z)$ closely follows the neutral hydrogen fraction, $x_{\rm HeI}(z)=x_{\rm HI}(z)$. The doubly ionized helium fraction becomes dominant ($x_{\rm HeIII}\gtrsim0.5$) at $z\lesssim3$. Consequently, in the redshift range $3\lesssim z\lesssim6$ the singly ionized helium fraction is the most abundant. We note that increasing $\eta$ from 0.1 to 1 does not significantly alter the relative abundances of the helium ionization states, since $x_{\rm HI}(z)\ll1$ at $z\lesssim6$. Therefore, spectral features caused by HeI and HeII absorption are expected to be present in the continuum spectra of distant sources at $z\gtrsim3$. In order to analyse these effects, in this section we estimate the optical depths of the intergalactic medium in the Lyman-series lines of HeI atoms and HeII ions.
\begin{table}
\centering
\caption{The wavelengths $\lambda^{\rm HeI}_{\rm 1-n}$ ({\AA}) and oscillator strengths $f^{\rm HeI}_{\rm 1-n}$ for the first 10 HeI lines of Lyman series \cite{Wiese2009}.}
\begin{tabular}{ccc|ccc }
\hline
\hline
1-n&$\lambda^{\rm HeI}_{\rm 1-n}$&$f^{\rm HeI}_{\rm 1-n}$&1-n&$\lambda^{\rm HeI}_{\rm 1-n}$&$f^{\rm HeI}_{\rm 1-n}$  \\
\hline
1-2& 591.412& 2.775$\cdot10^{-8}$&1-7&512.099& 0.0086306 \\
1-3& 584.334& 0.276250& 1-8&509.998& 0.0054073 \\
1-4& 537.030& 0.07346& 1-9&508.643& 0.0036108 \\
1-5& 522.213& 0.029873& 1-10&507.718& 0.0025304 \\
1-6& 515.617& 0.015045& 1-11&507.058& 0.0018420 \\ 
\hline
\hline
\end{tabular}
\label{tab3}
\end{table}

In the same way as for hydrogen, one can deduce the expression for optical depth in the lines of Lyman series of neutral helium atoms HeI,
\begin{eqnarray}
\tau_{\rm Ly_n}^{\rm HeI}&=& 2.2985\cdot10^{10}\lambda^{\rm HeI}_{\rm 1-n} f^{\rm HeI}_{\rm 1-n} Y_{\rm p}\frac{\Omega_{\rm b}h(1+z)^3x_{\rm HeI}(z)}{\sqrt{\Omega_{\rm m}(1+z)^3+\Omega_\Lambda}}, \nonumber\\
&&\mbox{where} \quad z=(1+z_{\rm s})\frac{\lambda^{\rm HeI}_{\rm 1-n}}{\lambda}-1.
\end{eqnarray}
The wavelengths and oscillator strengths for the first 10 Lyman lines of HeI taken from \cite{Wiese2009} are listed in Tab. \ref{tab3}. In the formulas here and below the wavelengths $\lambda^\alpha$, $\lambda^\beta$ ... are expressed in cm. The 1–2 transition is partially dipole-forbidden by quantum selection rules and therefore has a very small oscillator strength, resulting in negligible absorption. In the left panel of Fig.~\ref{tauLyHe}, we show the wavelength dependence of
$\tau^{\rm HeI}_\beta$ for sources at different redshifts, while the right panel
displays the total optical depth from the first ten HeI Lyman-series
lines, $\sum_{n=2}^{11}\tau^{\rm HeI}_n$.
 We find that $\tau^{\rm HeI}_\beta\gtrsim1$ and $\sum_{n=2}^{11}\tau^{\rm HeI}_n\gtrsim1$ at $\lambda\gtrsim4000$~{\AA} for sources at $z\gtrsim6$. At lower redshift $\sum_{n=2}^{11}\tau^{\rm HeI}_n\ll1$.

The optical depth in the Lyman continuum of HeI atoms is as follows
\begin{eqnarray}
\tau^{\rm HeI}_{\rm c} (\lambda,z_{\rm s})&=&\frac{cn^0_{\rm He}}{H_0}\int^{z_{\rm s}}_{z_{\rm o}}\frac{x_{\rm HeI}(z) (1+z)^2 \sigma^{\rm HeI}_{\rm c}(\lambda,z)dz}{\sqrt{\Omega_m(1+z)^3+\Omega_\Lambda}}, \nonumber\\
&&\mbox{where} \quad n^0_{\rm He}=Y_{\rm P}\frac{\Omega_{\rm b}}{m_{\rm He}}\frac{3H_0^2}{8\pi G}.
\label{tauHeLyc}
\end{eqnarray}
We use the approximation for effective cross-section of absorption in the HeI Lyman continuum proposed by Yan et al. \cite{Yan1998},
\begin{eqnarray}
\hskip-0.8cm\sigma^{\rm HeI}_{\rm c}(\lambda,z)&=&7.4\cdot10^{-18}\left[a_1x^{-s}+(1-a_1)x^{-(s+1)}\right] \nonumber\\
&+&\frac{7.33\cdot10^{-22}}{E(keV)^{7/2}}\left[1+a_2x^{-1/2}e^{-a_3x^{-1/2}}\right],
\end{eqnarray}
for $x=(E/24.58\,\mathrm eV)(1+z_{\rm o})/(1+z)\ge1$, where $E=ch_{\rm P}/\lambda$, and $\sigma^{\rm HeI}_{\rm c}=0$ for $x<1$. The values of best-fit parameters here are as follows: $a_1=7.3861$, $s=3.9119$, $a_2=-3.2491$, $a_3=1.1783$. The resulting cross-section is shown in the left panel of Fig. \ref{sigma-tauHeLyc} for the rest frame of Earth observer transferred from the absorber reference system at different redshifts.

In the right panel, we present the optical depths in HeI Lyman continuum for sources at different redshifts. We note a substantial difference between $\tau^{\rm HeI}_{\rm c}(\lambda,z_{\rm s})$ obtained for the two $x_{\rm HI}$ approximations, particularly for sources at $z_{\rm s}\simeq6-8$.

These results are obtained for $\eta=0.1$. When $\eta=1$, the optical depths slightly increase for sources at redshifts $z\lesssim5.5$ and remain nearly unchanged for sources at higher redshifts. Overall, the optical depth of the smooth intergalactic medium in the HeI Lyman continuum remains below unity for sources at $z\lesssim5.5$.

We compute the optical depths in the Lyman-series lines and continuum of singly ionized helium using hydrogen atomic data, since HeII is a hydrogen-like ion: $\lambda^{\rm HeII}_{ij}=\lambda^{\rm HI}_{ij}/4$, $f^{\rm HeII}_{ij}=f^{\rm HI}_{ij}\mu^{\rm HI}/\mu^{\rm HeII}$, $\mu^{\rm HI}/\mu^{\rm HeII}\approx0.9996$. The results are shown in Fig. \ref{tauHeII}. 

\begin{figure}
\includegraphics[width=0.49\textwidth]{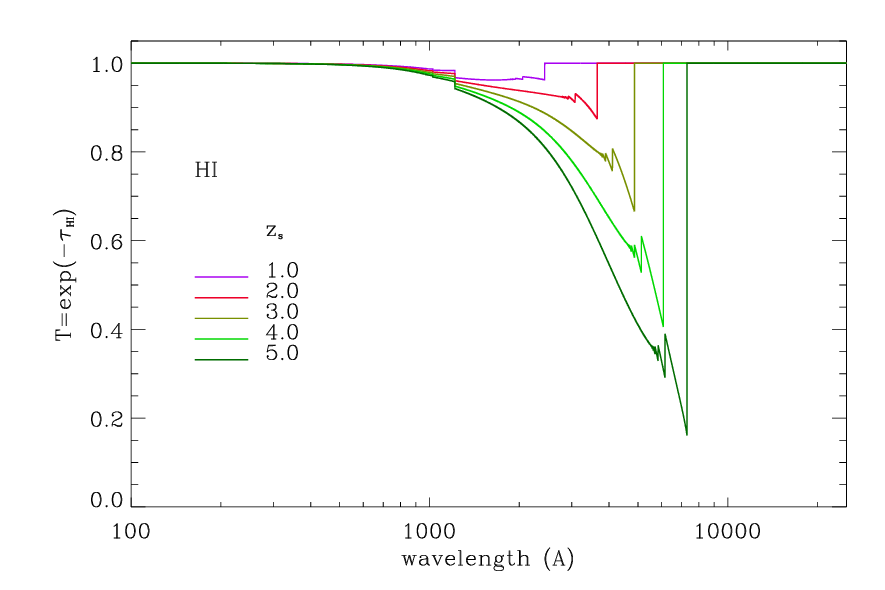}
\includegraphics[width=0.49\textwidth]{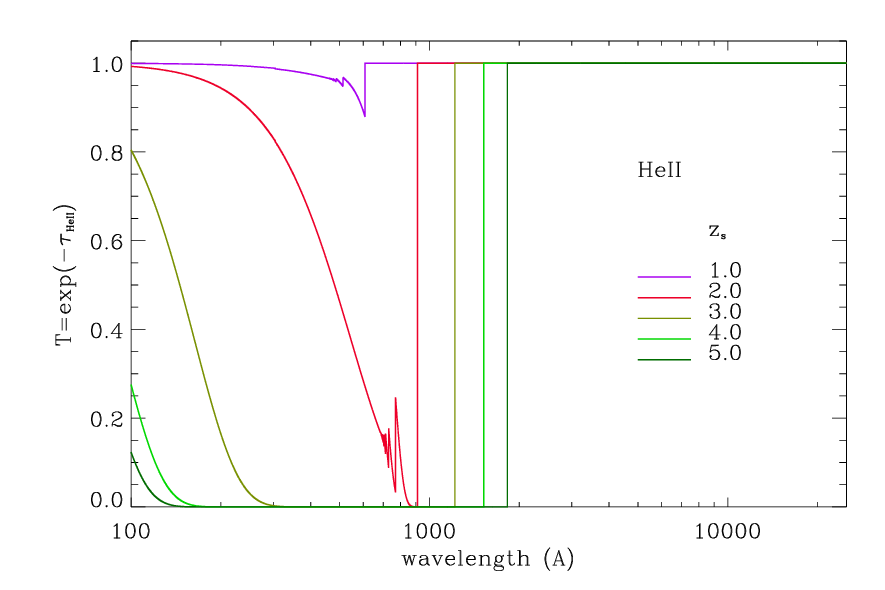}
\includegraphics[width=0.49\textwidth]{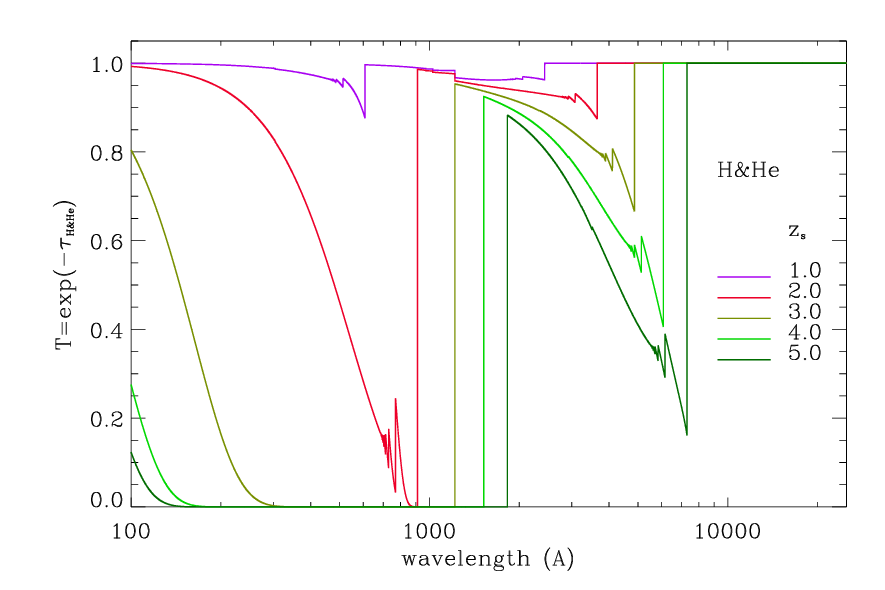}
\caption{Spectral transmittance functions for absorption by HI (left panel),  HeII (middle), and HI+HeI+HeII (right) for sources at $z_{\rm s}\in1-5$. }
\label{sptr1}
\end{figure}

\begin{figure*}
\includegraphics[width=0.49\textwidth]{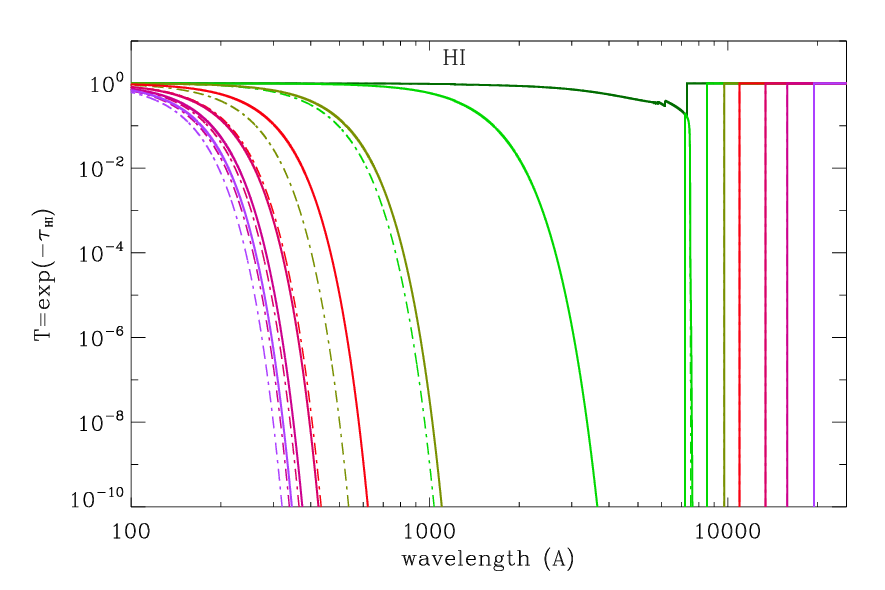}
\includegraphics[width=0.49\textwidth]{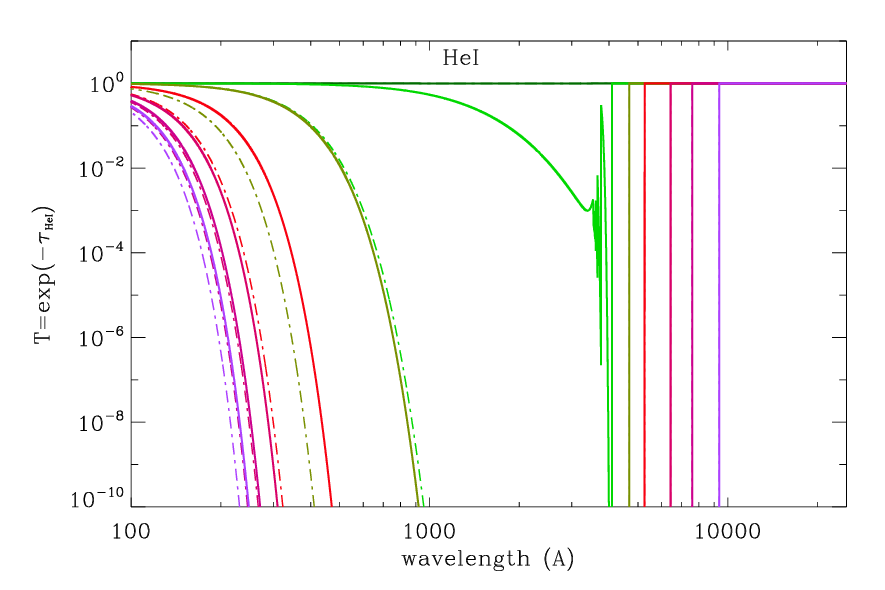}\\
\includegraphics[width=0.49\textwidth]{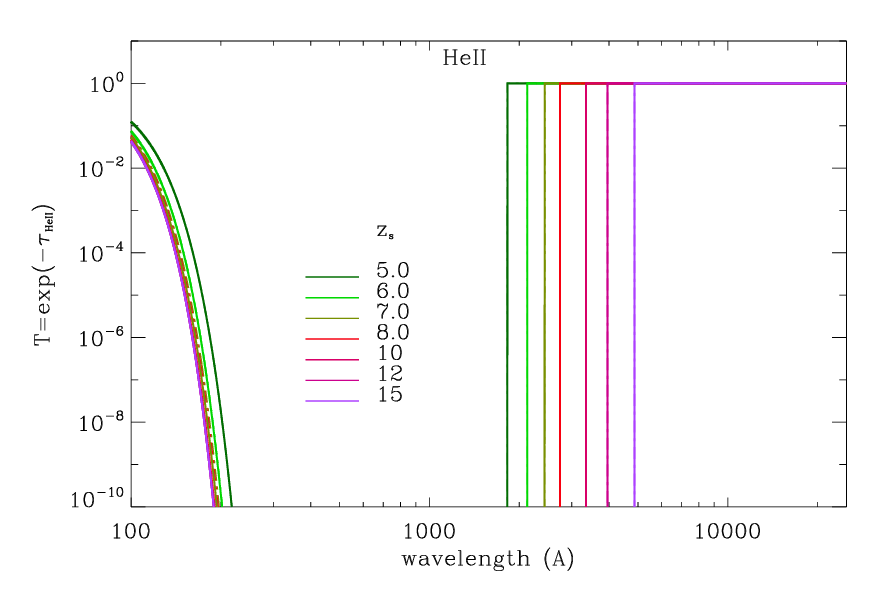}
\includegraphics[width=0.49\textwidth]{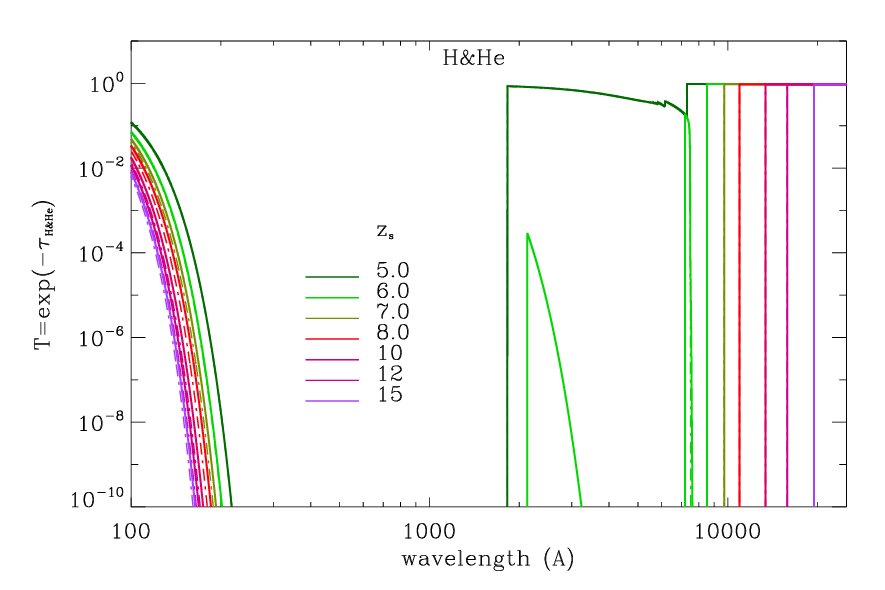}
\caption{Spectral transmittance functions for absorption by HI (upper left),  HeI (upper right), HeII (lower left) and HI+HeI+HeII (lower right) for sources at $z_{\rm s}\in5-15$. Dash-dotted curves correspond to model 1 of the $x_{\rm HI}$ approximation, solid curves to model 2, and colour indicates the source redshift.}
\label{sptr}
\end{figure*}

In the left panel, we present the optical depths in the first 39 lines of HeII Lyman series $\tau^{\rm HeII}_{\rm Ly}(\lambda)=\sum_2^{40}\tau^{\rm HeII}_{\rm Ly_n}$ for the sources at different redshifts. The differences of these dependences and corresponding HI results (Fig. \ref{tauLyH}, right panel) arise from differences of wavelengths  values, oscillator strengths and z-dependences of $x_{\rm HeII}$ and $x_{\rm HI}$, shown in Fig. \ref{xHI-xHe}. In the right panel the HeII Lyman-continuum optical depth $\tau^{\rm HeII}_{\rm Lyc}(\lambda)$ is presented for sources at $z_{\rm s}\in1-15$. Thus, absorption by HeII dominates the extreme ultraviolet part of the spectrum, producing a sharp short-wavelength cutoff that complements the long-wavelength Ly$\alpha$ absorption by HI.

\section{Spectral transmittance functions of intergalactic medium}

\begin{figure*}
\includegraphics[width=0.49\textwidth]{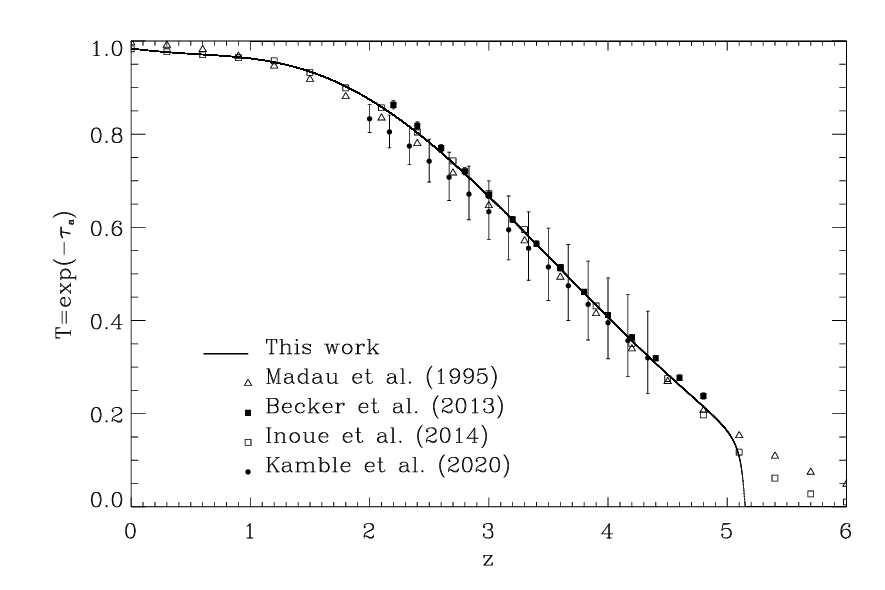}
\includegraphics[width=0.49\textwidth]{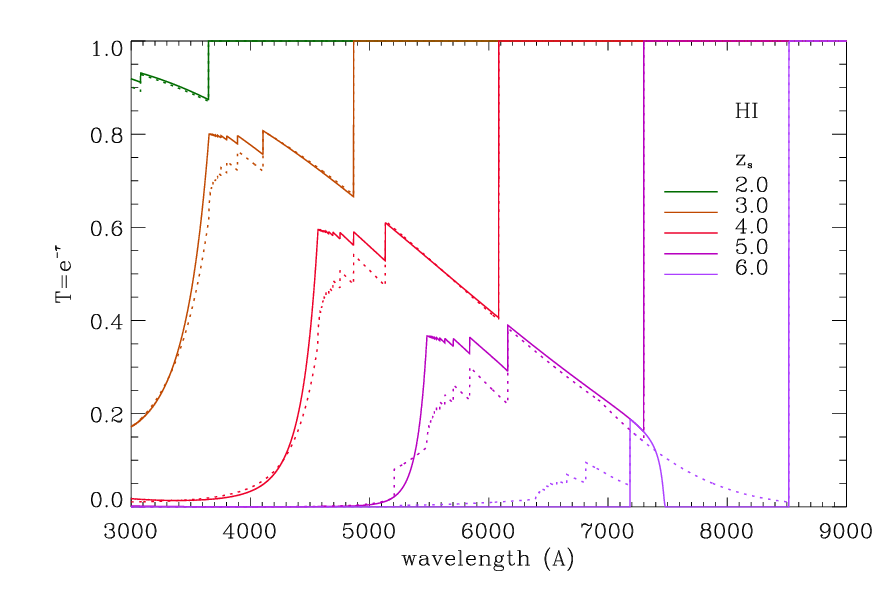}
\caption{Ly$\alpha$ transmission as a function of redshift (left panel) and hydrogen transmittance functions for sources at $z_{\rm s}\in2-6$ calculated with the fitting factor $\xi=27.5$ in equation (\ref{tauLyc}) (right panel). The dotted curves in the right panel show the analytical approximations of Inoue et al. (2014).}
\label{sptr2}
\end{figure*}

\begin{figure*}
\includegraphics[width=0.49\textwidth]{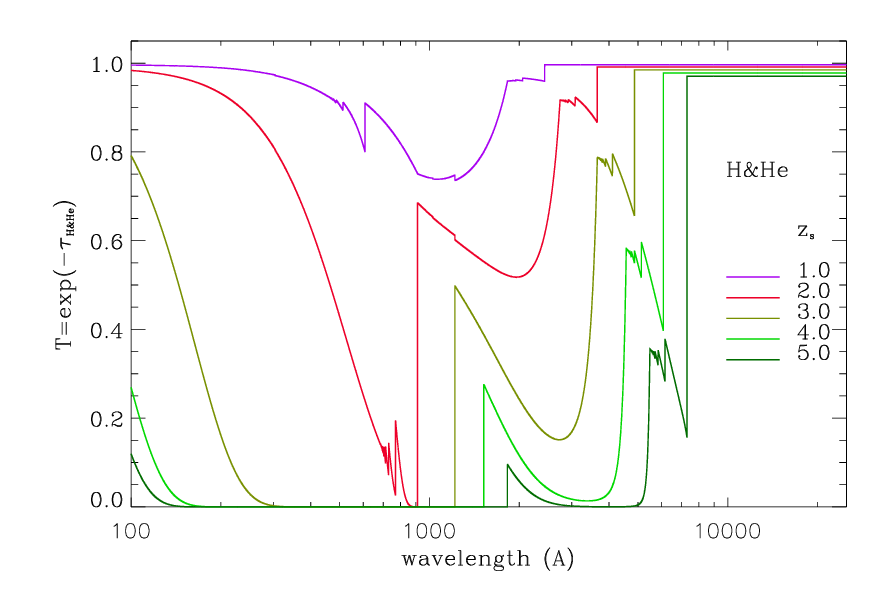}
\includegraphics[width=0.49\textwidth]{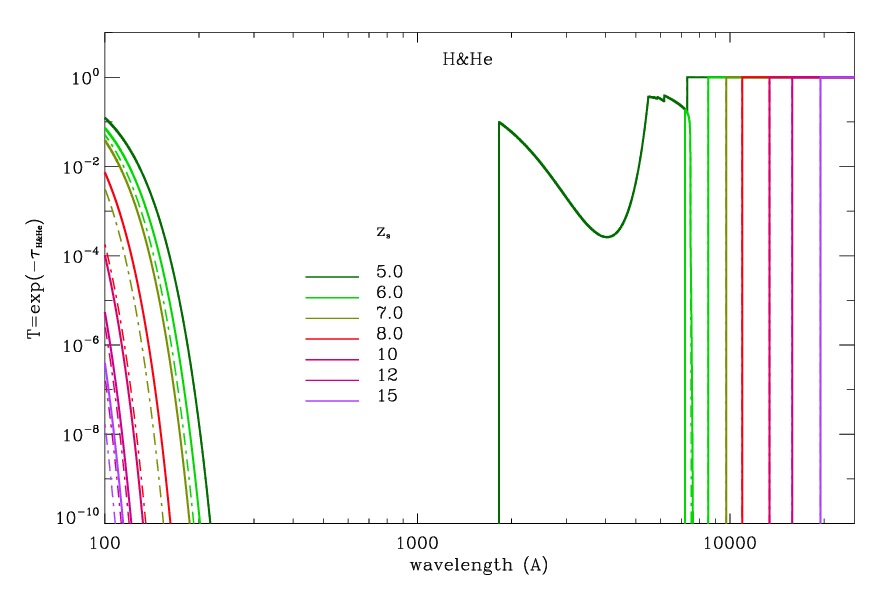}
\caption{Total spectral transmittance functions for sources at $z_{\rm s}\in1-5$ (left panel) and $z_{\rm s}\in5-15$ (right panel), calculated using the fitting factor $\xi=27.5$ in equation (\ref{tauLyc}).}
\label{sptrb}
\end{figure*}

In this and the following sections, we analyse how each absorption process described above, as well as their combined effect, modifies the continuum spectra of sources located at redshifts $1 \leq z_{\rm s} \leq 15$.  

Let $F_{\lambda^{({\rm o})}}(z_{\rm s})$ denote the observed spectral flux at wavelength $\lambda^{({\rm o})}$ from a source at redshift $z_{\rm s}$, including intergalactic absorption, and let $F_{\lambda^{({\rm o})}}^{(0)}(z_{\rm s})$ denote the corresponding flux in the absence of absorption. The spectral transmittance function is then defined as their ratio:
\begin{equation}
T(\lambda^{({\rm o})}; z_{\rm s}) \equiv \frac{F_{\lambda^{({\rm o})}}(z_{\rm s})}{F_{\lambda^{({\rm o})}}^{(0)}(z_{\rm s})} = \exp\left[-\tau(\lambda^{({\rm o})}; z_{\rm s})\right].
\end{equation}
It is independent of the intrinsic source spectrum. In the general case, the optical depth includes the contributions of all absorbers along the line of sight at the local wavelength $\lambda(z)=\lambda^{({\rm o})}(1+z_{\rm o})/(1+z)$. We also analyse the transmittance functions associated separately with HI, HeI, and HeII absorption.

In the log-norm Fig.~\ref{sptr1} we present the spectral transmittance functions of the intergalactic medium for absorption by H and He for $z_{\rm s} \in 1-5$. The HI absorption troughs deepen from approximately $\sim$0.02 to $\sim$0.2 as the source redshift increases from 1 to 5  (top panel). We omit the figure for HeI transmittance functions since they $\approx 1$ for all sources in this range because the number density of HeI atoms is small for $\eta\in0.1-1$ (see lower panel of Fig. \ref{xHI-xHe}). In this redshift range most of the helium atoms are in the singly and doubly ionized states and for $x_{\rm HeII}>0.1$ the number density of HeII ions is essentially larger than HI atoms and HeII ions are the main absorber at $\lambda < \lambda^{\rm HI}_{\alpha}(1+z_{\rm s})/4$. The transmittance functions presented in the middle and bottom panels of Fig.~\ref{sptr1} well illustrate that. One can see also that for sources at $z_{\rm s}\lesssim1$ intergalactic transmittance function $0.9\lesssim T\le 1$.

In the log-log Fig.~\ref{sptr} we present the spectral transmittance functions of the intergalactic medium for absorption by HI, HeI, HeII and total HI+HeI+HeII for $z_{\rm s} \in 5-15$. It illustrates the formation of deep absorption troughs in the spectra of distant sources caused by absorption in the Lyman-series lines and continua of HI, HeI and HeII, both individually and in combination. All absorbers contribute significantly to shaping the observable spectra of sources at $z \gtrsim 5$.
As expected, $T \approx 1$ at wavelengths redward of the HI Ly$\alpha$ break, $\lambda \gtrsim 1216(1+z_{\rm s})$~{\AA}. At sufficiently short wavelengths ($\lambda \lesssim 10$~{\AA}), the photoionization cross-sections rapidly decrease, so the IGM becomes increasingly transparent in this homogeneous approximation (not shown in figures).
The width and depth of the global absorption troughs increase with source redshift. The long-wavelength edge of the troughs is set by HI Ly$\alpha$ absorption, whereas the gradual short-wavelength edge is mainly controlled by the decreasing He II Lyman-continuum opacity. At a transmittance level of $\sim 10^{-3}$–$10^{-2}$, the two adopted reionization histories produce different amounts of weak residual transmission within the broad absorption trough at $z \simeq 5.5-6$. In the homogeneous limiting case considered here, this illustrates the sensitivity of the transmission to the adopted mean ionization history. In any case, these calculations show where opacity becomes saturated and where residual EUV transmission survives in the homogeneous limit.

We now compare the predictions of our analytical approach with observational data and with other approaches based on the statistical distribution of intergalactic absorbers. The most accurate measurements were made in the optical region of the spectrum to sources at $z \leq 6$ \cite{Becker2013,Inoue2014,Kamble2020}. To compare with them we compute the hydrogen transmission functions for sources at redshift $z = 2, 3, 4, 5,$ and 6. In the left panel of Fig.~\ref{sptr2} we present our Ly$\alpha$ transmission as a function of redshift (solid lines), together with the transmission functions derived from observational data. We find good agreement at $z \leq 5$. The steep decline of our transmittance function at $z > 5$ reflects the late reionization scenario inferred from the Kageura+Bosman and Glazer+Bosman data sets. At lower redshifts, the transmission curve derived in \cite{Inoue2014} exhibits a noticeably shallower slope at $z \lesssim 1.2$, consistent with the behaviour of $x_{\rm HI}(z)$ shown in the top panel of Fig.~\ref{xHI-xHe}. This figure provides a direct comparison between our model and observationally inferred transmission functions.

Comparison of our results presented in the top panel of Fig.~\ref{sptr1} with Fig. 4 in \cite{Inoue2014} shows good agreement for Ly$\alpha$–Ly$\beta$ transmission for sources at $z_{\rm s} \leq 5$, but a significantly different behaviour at shorter wavelengths and continuum. Our calculation refers to the optical depth of the smoothly distributed diffuse intergalactic medium with prescribed ionization fractions $x_{\rm HI}(z)$, $x_{\rm HeI}(z)$, and $x_{\rm HeII}(z)$. It therefore differs conceptually from the mean transmission functions of \cite{Madau1995} and \cite{Inoue2014}, which are dominated by the statistical distribution of discrete absorbers such as Ly$\alpha$-forest clouds, Lyman-limit systems, and damped Ly$\alpha$ systems. In the smooth-IGM case, the rapid decline of the photoionization cross-section with frequency, $\sigma_\nu \propto \nu^{-3}$, can dominate over the increase of the absorbing path length, so that the continuum optical depth decreases towards shorter wavelengths. In the formalism of \cite{Madau1995} and \cite{Inoue2014}, the effective optical depth is averaged over the absorber population and involves the factor $1 - \exp(-N_{\rm HI}\sigma_\nu)$, which saturates for optically thick systems. As a result, the contribution of Lyman-limit systems becomes insensitive to the exact frequency dependence of the cross-section once $\tau_{\rm abs} \gtrsim 1$. At the same time, shorter observed wavelengths correspond to a larger redshift interval over which photons remain above the Lyman edge, increasing the effective path length for absorption (see Fig~\ref{scheme}). Since the Lyman-continuum opacity is dominated by these optically thick absorbers, the cumulative number of absorbers increases towards shorter wavelengths. Consequently, the effective optical depth increases with frequency despite the decrease of the microscopic cross-section in some finite ranges of $\lambda$.

In our approach, a Lyman-continuum transmittance similar to that of \cite{Inoue2014} can be obtained by introducing the phenomenological factor $\xi$ in equation (\ref{tauLyc}). We find that equation (\ref{tauLyc}) with $\xi=27.5$ well reproduces the wavelength dependence of the Lyman-continuum transmittance predicted by Inoue et al. \cite{Inoue2014}, as illustrated in the right panel of Fig.~\ref{sptr2}. Thus, it can be treated as a phenomenological opacity-enhancement parameter that quantifies the factor by which the effective opacity of a clumpy absorber population can exceed that of the homogeneous medium in the Lyman-limit range. The dotted lines are the linear approximations (21)-(29) in \cite{Inoue2014}. This figure demonstrates good agreement in both the Ly$\alpha$ and Lyman-continuum spectral ranges at  $z\le5$. In Fig.~\ref{sptrb} we show the total spectral transmittance functions of a hydrogen–helium medium for sources at $z_{\rm s} \in1-5$ (left panel) and $z_{\rm s} \in 5-15$ (right panel) for this model. We note significant changes in $T(\lambda,z_{\rm s})$ in the Lyman-continuum spectral range for sources at $z_{\rm s}=1-5$ (see the bottom panel of Fig.~\ref{sptr1} for comparison), but practically no change for sources at $z_{\rm s}>5$, because the broad absorption region is already dominated by HI Ly$\alpha$ absorption.

If one treats the Lyman-continuum transmittance functions shown in Fig. 4 of \cite{Inoue2014} as observationally supported, then our approach can be used to approximate them analytically. In this case, the fitting factor $\xi$ in equation (\ref{tauLyc}) quantifies how much the continuous absorption along the line of sight in a clumped intergalactic medium exceeds that in a homogeneous medium with the same Ly$\alpha$-absorption.

All computations presented above in tables and figures are done using our code \texttt{rtf\_H-He.f90} available at GitHub with link \url{https://github.com/luspav/spectral-transfer-function-hhe}.

\section{Illustrative applications to source spectra}

To illustrate the impact of intergalactic hydrogen and helium absorption on the spectra of sources at different redshifts, we use toy continuum spectra of starless halos and a more realistic spectral energy distribution of a low-metallicity dwarf galaxy in their respective rest frames.

We model the spectral flux from sources at redshifts $z_{\rm s}\in0.02-15$ as observed by an Earth-based observer, taking into account the absorption by the intergalactic medium, luminosity distance, cosmological expansion, and Thomson scattering by free electrons. Absorption by the interstellar medium of the Milky Way, which is important but well studied, is neglected here. 

The spectral luminosities of the first sources of light formed during the Cosmic Dawn are poorly constrained. The commonly accepted hypothesis is that the first stars and galaxies formed in dark matter halos. Recently, however, the Hubble Space Telescope confirmed the existence of a new class of objects -- galaxies without stars \cite{Anand2025}, originally discovered in a radio survey with the Five-hundred-meter Aperture Spherical Telescope \cite{Zhou2023}.
The discovered object, dubbed Cloud~9, is a halo of hot baryonic gas embedded in a dark matter halo, in which the conditions required for star formation never developed. Its estimated distance is 4.4 Mpc, corresponding to $z=0.001$ in the standard $\Lambda$CDM model \cite{Anand2025}. It is plausible that similar objects existed in significant numbers in the early Universe, making them a convenient and physically simple class of sources for modelling continuous emission spectra in their rest frames.
\begin{figure}
\includegraphics[width=0.49\textwidth]{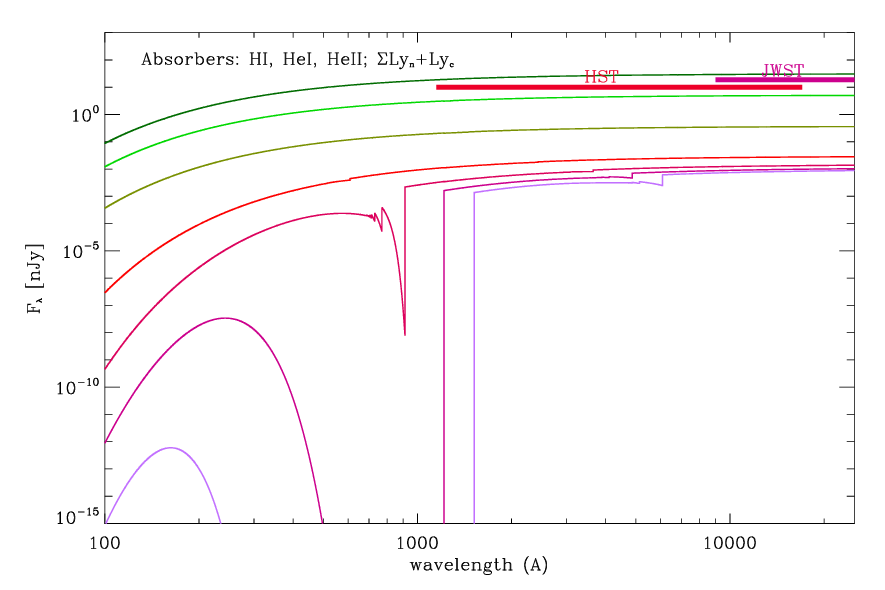}
\includegraphics[width=0.49\textwidth]{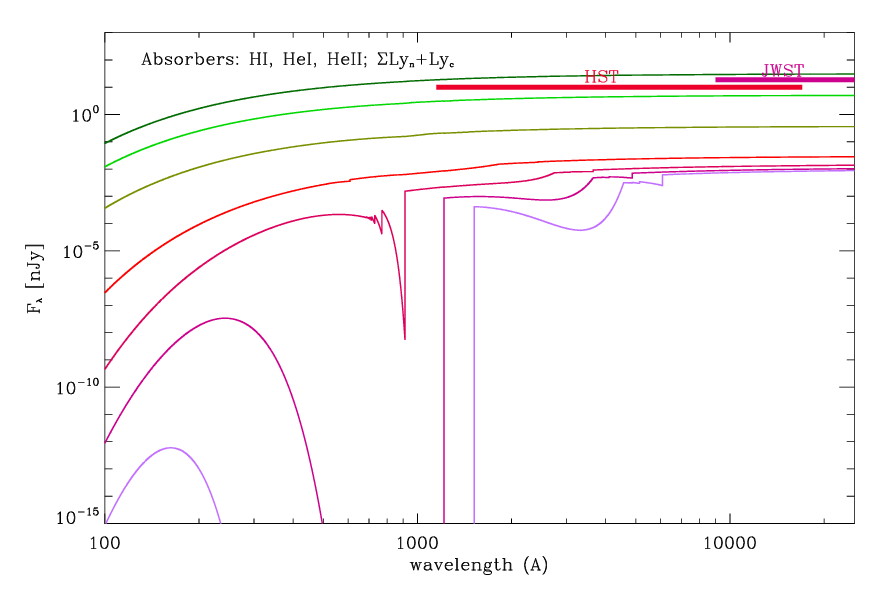}
\caption{Spectral fluxes $F_\nu$ in the HST-JWST spectral ranges from halos at $z_{\rm s}=0.02,\,0.05,\,0.2,\,1.0,\,2.0,\,3.0,\,4.0$ (from top to bottom) in the model with $\xi=1$ (top panel) and $\xi=27.5$ (bottom panel). }
\label{sflux1}
\end{figure}

In our calculation, we assume that such sources are virialized spherical halos with virial temperatures  \cite{Barkana2001,Bromm2011},
\begin{equation}
T_{\rm vir}=2\cdot10^4\left(\frac{\mu_{\rm H}}{1.2}\right)\left(\frac{M_{\rm h}}{10^8\,\mathrm{M_{\odot}}}\right)^{\frac{2}{3}}\left(\frac{\Delta_{\rm vir}}{178}\right)^{\frac{1}{3}}\left(\frac{z_{\rm vir}+1}{10}\right)\,\mathrm{K},
\label{Tvir}     
\end{equation}
where $\mu_{\rm H}$ is mass per H atom, $z_{\rm vir}$ is redshift of halo virialization, and $\Delta_{\rm vir}$ is density contrast at the moment of virialization. The halo mass $M_{\rm h}/M_{\odot}$ and its physical virial radius $R_{\rm h}/{\rm kpc}$ after virialization are related as \cite{Novosyadlyj2022}
\begin{equation}
R_h = 1.5 \left( \frac{M_h}{10^8 M_\odot} \frac{178}{\Delta_{\rm vir}} \frac{0.143}{\Omega_m h^2} \right)^{1/3} \frac{10}{1+z_{\rm vir}} \quad {\rm kpc}. 
\label{Rh}
\end{equation}
These relations indicate that, within the standard $\Lambda$CDM model, halos with masses $M_{\rm h}\sim10^9$--$10^{10}\,M_\odot$ at redshifts $z=5-15$ have virial temperatures in the range $(2-7)\cdot10^5$~K and radii of $2-12$~kpc. Kageura et al. (2025) assume that  massive halos could be the primary sources of reionization. 

To isolate the effect of intergalactic absorption on the emergent spectra, we assume that halos at different redshifts share the same characteristic temperature, $T_{\rm vir}=2.5\cdot10^5$~K, and for simplicity set $z_{\rm vir}=z_{\rm s}$.

We assume that the continuum emission is dominated by free–free and free–bound processes. Recent refinements of the classical free–free emission formalism based on high-precision quantum mechanical calculations provide updated Gaunt factors and radiative coefficients over a wide range of plasma conditions \cite{Hoof2014,Chluba2020,Pradler2021}. Our approach follows the same physical principles but applies them to a different cosmological regime, emphasizing analytic transparency.

The angle-integrated emissivity of an optically thin plasma in the source rest frame is given in \cite{Lang1974,Spitzer1978,Rybicki1979,Draine2011} as follows 
\begin{equation}
\epsilon_\nu= 6.8\cdot10^{-38}\frac{Z^2 (g_{\rm fb}+g_{\rm ff}) n_{\rm e} n_{\rm i}}{\sqrt{T_{\rm vir}}\exp{\left(-h_{\rm P}\nu/k_{\rm B}T_{\rm vir}\right)}} \,\, {\rm \frac{erg}{s\cdot cm^3Hz}},
\end{equation}
where $Z$ is mean charge of particles in plasma, $g_{\rm fb}$ and $g_{\rm ff}$ are Gaunt factors for free-bound and free-free transitions. For completely ionized hydrogen-helium plasma $Z=1+Y_{\rm p}/(4(1-Y_{\rm p}))\approx1.07$, product of number density of electrons and ions for such plasma $n_{\rm e} n_{\rm i}\approx1.26n^2_{\rm H}$. This approximation assumes a fully ionized hydrogen–helium plasma. The emitting gas mass is taken to be the baryonic fraction, $M_{\rm gas}=(\Omega_b/\Omega_m)M_h$. In the frequency and temperature ranges considered here, $g_{\rm fb}\approx1$ and $g_{\rm ff}\approx1.2$ \cite{Rybicki1979}. The spectral luminosity in the rest frame of source is $L_{\nu_{\rm s}}=\frac{4}{3}\pi R_{\rm h}^3\epsilon_{\nu_{\rm s}}$. The observed spectral flux is related to the emitted spectral luminosity through the luminosity distance (e.g. \cite{Peebles1993}, p. 330) as follows
\begin{equation}
F_{\nu^{\rm (o)}}(z_{\rm s}) =\frac{1+z_{\rm s}}{1+z_{\rm o}} \frac{L_{\nu^{\rm (s)}}}{4\pi D_L^2} \exp\left[-\tau(\nu^{\rm (o)}; z_{\rm s})\right],
\label{flux2}
\end{equation}
where $\nu^{\rm (s)} = \nu^{\rm (o)}(1+z_{\rm s})/(1+z_{\rm o})$ and $D_L(z_{\rm s})$ is the luminosity distance to the source  
\begin{equation}
D_L(z_{\rm s}) = \frac{(1+z_{\rm s})c}{(1+z_{\rm o})H_0} \int_{z_{\rm o}}^{z_{\rm s}} \frac{dz}{\sqrt{\Omega_m(1+z)^3 + \Omega_\Lambda}}.
\end{equation}

\begin{figure}
\includegraphics[width=0.49\textwidth]{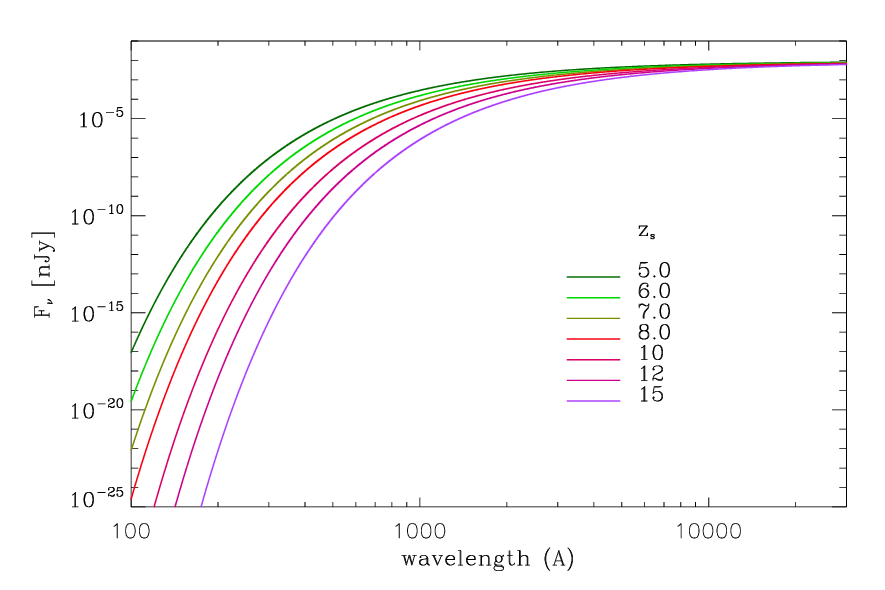}
\includegraphics[width=0.49\textwidth]{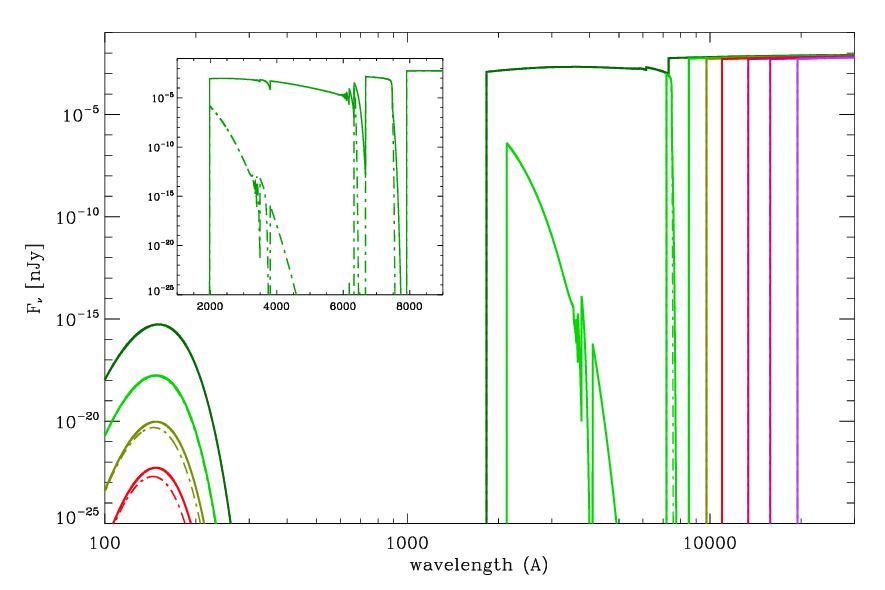}
\includegraphics[width=0.49\textwidth]{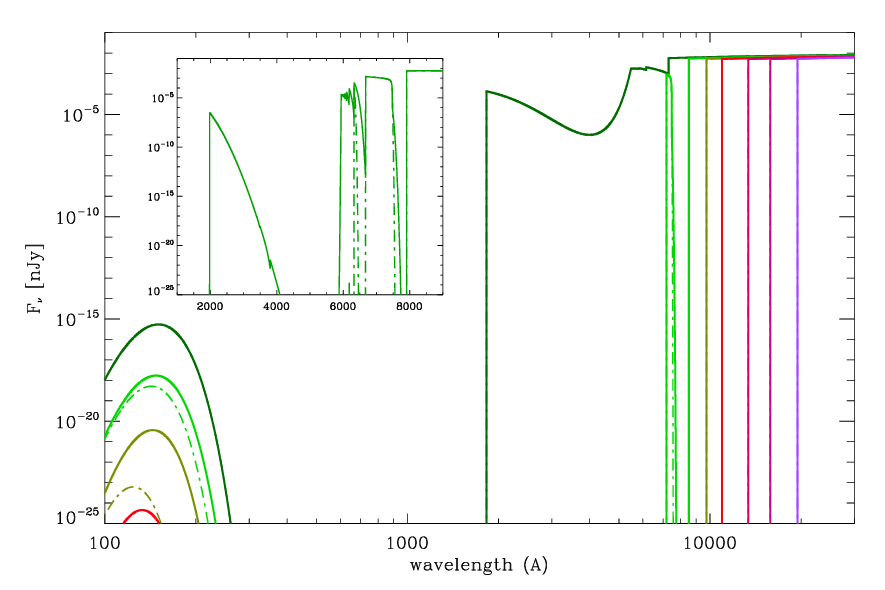}
\caption{Spectral fluxes observed at Earth from haloes at $z_{\rm s}\in5-15$ without intergalactic absorption (top panel) and including absorption by HI, HeI, and HeII, as well as Thomson scattering (middle panel for case $\xi=1$, bottom panel for case $\xi=27.5$).}
\label{sflux2}
\end{figure}

This is not intended as a detailed model of the spectral luminosity of a RELHIC; it is used solely as a smooth continuum test spectrum to illustrate the effect of the derived transmittance functions. Cosmological hydrodynamical simulations of starless halos, termed REionization-Limited HI Clouds (RELHICs), were performed by Ben\'itez-Llambay et al. \cite{Benitez2017}, who predicted their thermodynamic properties, gas-density profiles, and temperature profiles in detail. Within our simplified analytical model, we estimate the spectral fluxes from such objects placed at different redshifts and use the achieved observational sensitivities in the IR range for comparison.

 In Fig. \ref{sflux1}, we present the fluxes (in nJy) from Cloud9-like halos at redshift $z_{\rm s}\in0.02-4$ in the spectral range 900-30000 {\AA} covering the HST-JWST spectral ranges. The sensitivity achieved by HST is $\approx10$ nJy ($m_{\rm AB}\approx29$) \cite{Sankrit2025}. According to JWST documentation\footnote{https://jwst-docs.stsci.edu/jwst-near-infrared-camera/nircam-performance/nircam-sensitivity}, the Near Infrared Spectrograph (NIRSpec) covers the wavelength range $0.6$--$5.3\,\mu$m and can detect a flux of $\sim18.9$~nJy at $\lambda\simeq3\,\mu$m with a signal-to-noise ratio of 10 for an integration time of 10\,000~s. The HST and JWST spectral ranges and achieved sensitivities are shown by horizontal lines. This estimation shows that such starless clouds may be detectable by HST or JWST with the longest exposures at redshift $z\lesssim 0.05$. Intergalactic absorption features become noticeable in the continuum spectra of such clouds at $z\gtrsim1$. The spectral feature visible in Fig. \ref{sflux1} is caused by Ly$\alpha$ absorption, but the corresponding fluxes are far below the HST and JWST detection limits. 

The spectral fluxes in the wavelength range $100\le\lambda\le30\,000$~{\AA}\ from the Cloud9-like halos with mass $\sim 10^9$ M$_\odot$ and virialized temperature at redshifts $5\le z\le15$ are shown in Fig.~\ref{sflux2} to illustrate the broad and deep absorption troughs. The top panel displays the redshifted intrinsic spectra attenuated only by Thomson scattering, while the next two panels show the spectra transformed also by intergalactic absorption for $\xi=1$ (middle panel) and $\xi=27.5$ (bottom panel), respectively. As a result, the spectra of sources at $z_{\rm s}\in5-15$ exhibit wide and deep absorption troughs caused by H and He absorption. In the spectra of sources at $z\approx5$ we find significant absorption in the redshifted HI Ly$\alpha$ and Ly$\beta$ lines together with a broad flux depression extending to the redshifted HeII Ly$\alpha$ line. At $\lambda\le1824$~{\AA}, a broad and deep trough is produced by HeII Lyman-series and continuum absorption. In the spectra of sources at $z\sim5.5$ (inset in Fig.~\ref{sflux2}) more absorption spectral features are present. The solid curve (GBBI ionization history) shows HI Ly$\alpha$, Ly$\beta$, Ly$\gamma$, and other Lyman-series absorption lines, HeI Ly$\beta$ and Ly$\gamma$, and a wide and deep trough caused by HeII Lyman-series and continuum absorption. The dash-dotted curve (KBBI ionization history) shows substantially stronger absorption features because $x_{\rm HI}$ and $x_{\rm HeI}$ are larger in this case. The same absorption features become even stronger in the spectra of sources at $z\simeq6$. Within the amplitude range shown in Fig.~\ref{sflux2}, the spectra of sources at $z>6$ exhibit wide and deep troughs. Their long-wavelength edges are set by HI Ly$\alpha$ absorption at $\approx1216(1+z_{\rm s})$, whereas the short-wavelength edge is set by HeII Lyman-continuum absorption.

The predicted fluxes from such halos at $z_{\rm s}\in5-15$ therefore are well below the sensitivity limits of JWST. However, the cumulative contribution of a population of such halos, as well as possible enhancements due to additional physical processes (e.g., shocks, non-equilibrium ionization, or deviations from uniform density), may increase the observable signal and impact of their radiation on the state of intergalactic gas in the early Universe.

Other sources that may affect the thermal and ionization state of the intergalactic gas are dwarf galaxies (DGs) \cite{Atek2024, Izotov2024}. Since they are the most numerous sources of radiation during the Cosmic Dawn and Reionization epochs, their impact on the physical state of gas in halos and the intergalactic medium may be crucial.

\begin{figure}
\includegraphics[width=0.49\textwidth]{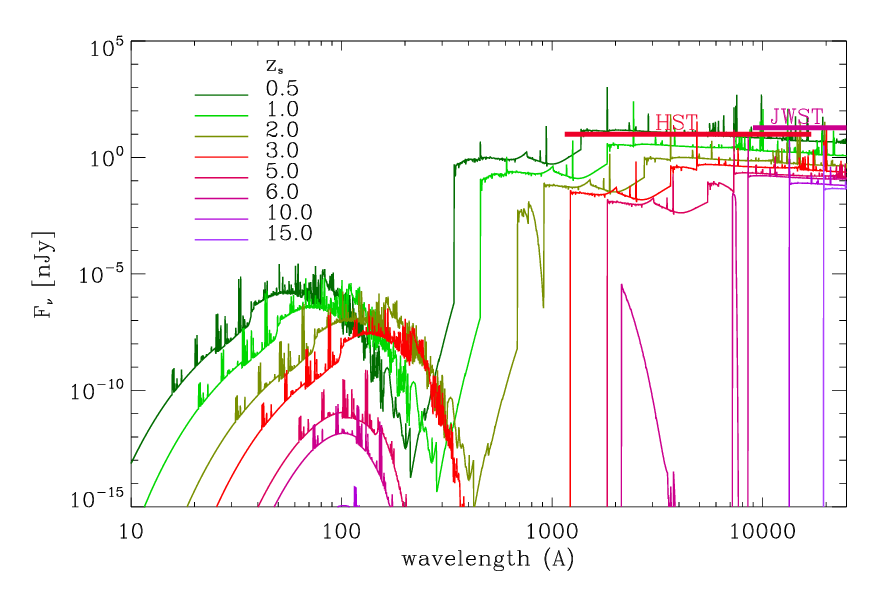}
\includegraphics[width=0.49\textwidth]{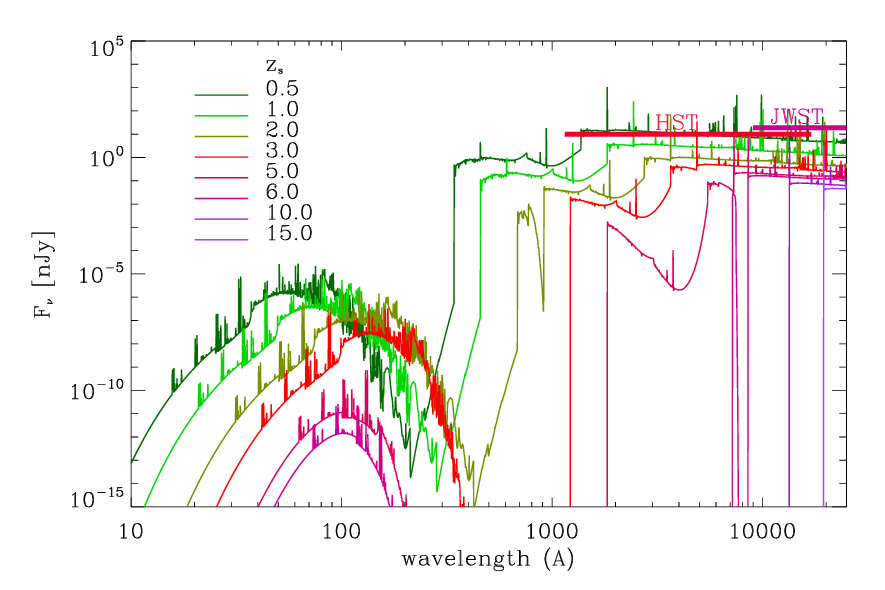}
\caption{Spectral fluxes of a dwarf galaxy with a total mass of $10^9$ M$_\odot$, placed at different redshifts, for two treatments of Lyman-continuum absorption: $\xi=1$ (top panel) and $\xi=27.5$ (bottom panel).}
\label{sdgflux}
\end{figure}

In Fig. \ref{sdgflux}, we present the spectrum of a low-metallicity dwarf galaxy (DG) with a total mass of $10^9\ M_\odot$ and active star formation, generated using the multicomponent photoionization modelling technique developed by Melekh et al. \cite{Melekh2025}. This technique models the radiative transfer of radiation from the central star-forming region through nebular gas evolving due to the expansion of a superwind bubble. The approach is based on the photoionization code CLOUDY v23.01 \cite{Gunasekera2023}, as well as on the hydrodynamical models of Weaver et al. \cite{Weaver1977} and Chevalier \& Clegg \cite{Chevalier1985}. The Illustris TNG50 simulation results \cite{Pillepich2019,Nelson2019} together with the BPASS evolutionary population synthesis models \cite{Eldridge2017} were used to obtain the input data required for this modelling. The code output is the intrinsic spectral luminosity expressed as $\nu_{\rm s}L_{\nu_{\rm s}}$. 

The spectra shown in Fig.~\ref{sdgflux} are obtained from the intrinsic spectrum by applying cosmological redshifting, intergalactic absorption, and Thomson scattering by free electrons. Our estimates show that the observed spectral fluxes from such sources are substantially higher than from Cloud9-like halos in the wavelength range accessible to HST and JWST, as well as at shorter wavelengths that are important for the heating and ionization of baryonic matter during the Cosmic Dawn and Reionization epochs.

Since for sources at $z_{\rm s}\lesssim1$ transmittance functions are in the range $\sim0.9-1$, the spectrum of DG at $z_{\rm s}=0.5$ in Fig. \ref{sdgflux} is practically its intrinsic spectrum with a large number of emission lines and two absorption features, the deep trough HeII Ly$_{\rm c}$ 228~{\AA} and shallow trough HI Ly$_{\rm c}$ 912~{\AA}, redshifted by a factor of 1.5. The asymmetry of the short-wavelength wing of the Ly$\alpha$ 1216~{\AA} line, barely noticeable in Fig.~\ref{sdgflux} in the spectra of sources at $z_{\rm s}\sim0.5-1$, increases to $z_{\rm s}\sim5$ and develops into a deep, dominant absorption trough for sources at $z_{\rm s}>6$. We see a similar trend in the HeII Ly$\alpha$ line 304~{\AA} for sources at $z_{\rm s}\lesssim5$. At $z_{\rm s}=5$, the spectrum exhibits a deep HeII Ly$\alpha$ trough, an HI Ly$\alpha$ trough of order unity, and an HI Lyman-continuum trough whose depth depends on $\xi$. In the spectrum of DG at $z_{\rm s}=6$, one can see the deep troughs HeII Ly$\alpha$, HI Ly$\beta$ and HI Ly$_{\rm \alpha}$ (from left to right) in the top panel whereas only the latter two are prominent in the bottom panel. The HI Ly$\alpha$ troughs are the dominant features in the spectra of sources at $z_{\rm s}>6$ (see also right bottom panel in Fig.~\ref{sptr} and right panel in Fig.~\ref{sptrb}).

\section{Conclusions}

We have investigated the impact of absorption by the diffuse intergalactic medium on the continuum spectra of sources at high redshifts $z\in1-15$. For that, we have constructed two self-consistent analytic approximations for the neutral hydrogen fraction, $x_{\rm HI}(z)$, and the helium ionization fractions, $x_{\rm HeI}(z)$, $x_{\rm HeII}(z)$, and $x_{\rm HeIII}(z)$, that are consistent with current constraints inferred from quasar spectra, galaxy surveys, and CMB polarization measurements. These approximations describe observationally motivated early \cite{Glazer2018,Bosman2022} and late \cite{Kageura2025,Bosman2022} reionization scenarios. At redshifts $z\le5$, the redshift dependence of $x_{\rm HI}$ has been deduced from the observationally inferred Ly$\alpha$ transmission in \cite{Becker2013,Inoue2014}. At $z\gtrsim5$, these ionization histories are used as an effective homogeneous representation of the globally averaged reionization history, so that the corresponding transmittance functions should be interpreted as a homogeneous limiting case.

For these ionization histories, we analyse the formation of absorption features in the continuum spectra of sources at $1\leq z_{\rm s}\leq15$. Assuming that neutral hydrogen and helium in a homogeneous diffuse intergalactic medium reside predominantly in their ground states, we compute the wavelength-dependent optical depths for the first 39 Lyman-series lines of HI and HeII, the first 10 lines of HeI, and the corresponding continua. For sources at $z_{\rm s}\lesssim 5$ $\tau_{Ly_{\rm \alpha}}^{\rm HI}\lesssim1$, whereas for sources at $z_{\rm s}>6$ it is $\gg1$ and decreases towards shorter wavelengths. The optical depths in the higher Lyman-series  lines and continuum show similar behaviour, but also depend on the adopted model of the medium (homogeneous or clumped). The optical depths in the HeI Lyman series lines and continuum for sources at $z_{\rm s}\lesssim 5$ are $\ll1$ and $\gtrsim1$ for $z_{\rm s}\gtrsim 6$. Remarkably, the optical depths in the HeII Lyman-series lines exceed unity already for $z_{\rm s}\gtrsim 1.5-2$. The constructed transmittance functions for sources at different redshifts, $T(\lambda;z_{\rm s})$, illustrate the formation of troughs in the continuum spectra of sources at different redshifts. For sources at $z_{\rm s}\le6$ distinct troughs are produced by HI Ly$\alpha$, HI Ly$\beta$ and HeII Ly$\alpha$. For sources at $z_{\rm s}>6$ the deep and broad trough produced by HI Ly$\alpha$ becomes dominant and encompasses the other absorption features. It has an abrupt long-wavelength edge at $\lambda=1216(1+z_{\rm s})$~{\AA} and a gradual short-wavelength edge mainly determined by the decreasing HeII Lyman-continuum opacity.

Overall, our results emphasize that absorption by hydrogen and helium in the intergalactic medium must be treated jointly when interpreting the spectra of high-redshift sources. The spectral transmittance functions derived here provide a compact and physically motivated tool for incorporating these effects into models of early cosmic radiation backgrounds and for assessing the imprint of reionization histories on observable spectra. Future observations with higher sensitivity may allow these features to be used as indirect probes of the ionization history of hydrogen and helium.

The distinctive feature of this work is that, in the framework of a homogeneous intergalactic medium with observationally constrained ionization histories, the Lyman-continuum transmission increases toward shorter wavelengths due to the $(\nu/\nu_{\rm pi})^{-3}$ dependence of the photoionization cross-section. This behaviour differs from that predicted by absorber-based models (e.g., \cite{Madau1995,Inoue2014}), in which the effective optical depth increases over a spectral range close to the ionization threshold owing to the cumulative contribution of optically thick systems, but eventually decreases at large $\nu/\nu_{\rm pi}\gg1$. It is explained as follows.  At a fixed observed wavelength, Ly$\alpha$ absorption occurs only within a narrow redshift interval around the resonance point, whereas ionizing photons accumulate continuum opacity over the longer part of their trajectory for which their local energy remains above the HI ionization threshold. In addition, the effective Ly$_c$ opacity is particularly sensitive to self-shielded systems with large HI column densities, which are not represented by a homogeneous mean neutral fraction. We have also illustrated how this discrepancy can be accounted for phenomenologically by introducing the fitting parameter $\xi$, which represents the enhancement of the effective opacity along the line of sight in the Lyman continuum and higher Lyman-series lines.

As an illustrative application, we modelled the spectral fluxes from starless gas halos with mass $\sim10^9$~M$_\odot$ and virial temperature $\sim2.5\cdot10^5$~K placed at different redshifts in the range $0.02\le z\le15$, assuming continuum emission dominated by free--free and free--bound processes. The resulting spectra clearly demonstrate how intergalactic absorption reshapes the intrinsic emission and produces characteristic broad absorption features in the continuum spectra of distant sources. The predicted fluxes of high-redshift halos, for which intergalactic absorption features become prominent, lie well below the sensitivity limits achieved by HST and JWST. We also estimated the spectral fluxes of an actively star-forming dwarf galaxy placed at different redshifts. The resulting spectra demonstrate that actively star-forming dwarf galaxies are potentially important contributors to the UV radiation field during reionization, although quantifying their global contribution requires a population model.


\section*{Acknowledgements}
This work was supported by the International Center of Future Science and the College of Physics of Jilin University (P.R. China).
BM acknowledges support from the National Research Foundation of Ukraine (grant 2023.03/0098).
The authors declare that an AI-based language model was used solely to assist with language editing and clarity of presentation.

\label{lastpage}
\end{document}